\documentclass[usegraphicx,usedcolumn,usenatbib]{mn2e}
\usepackage{graphicx}
\usepackage{txfonts}
\usepackage{rotating}
\usepackage{accents}
\usepackage{float}
\usepackage{epsfig}
\usepackage{epstopdf}
\usepackage{graphicx}
\usepackage{color}

\newcommand{\Msun}{${\rm M}_{\sun}$}
\bibpunct[,]{(}{)}{;}{a}{,}{,}

\title[Evolution of the Milky Way's oxygen gradient]{The time evolution of the Milky Way's oxygen abundance gradient}
\author[Moll\'{a} et al. ]{M. Moll\'{a}$^{1}$\thanks{E-mail:mercedes.molla@ciemat.es},
{\'A}.~I. D{\'\i}az$^{2,3}$, O. Cavichia$^{4}$, B.~K. Gibson$^{5,7}$, W.~J. Maciel$^{6}$,
 \newauthor
R.~D.~D. Costa$^{6}$, Y. Ascasibar$^{2,3}$, C.~G. Few$^{5,7}$  \\
$^{1}$ Departamento de Investigaci\'{o}n B\'{a}sica, CIEMAT, Avda. Complutense 40. E-28040 Madrid. (Spain)\\ 
$^{2}$ Departamento de F{\'\i}sica Te{\' o}rica, Universidad Aut{\'o}noma de Madrid, Cantoblanco, 28049 Madrid, Spain\\
$^{3}$ Astro-UAM, Unidad Asociada CSIC, Universidad Aut{\'o}noma de Madrid, 28049, Madrid, Spain\\
$^{4}$ Instituto de F{\'i}sica e Qu{\'i}mica, Universidade Federal de Itajub{\'a}, Av. BPS, 1303, 37500-903, Itajub{\'a}-MG, Brazil\\
$^{5}$ E.~A. Milne Centre for Astrophysics, University of Hull, Hull, HU6~7RX, United Kingdom\\
$^{6}$ Instituto de Astronomia, Geof{\'{\i}}sica e Ci{\^e}ncias Atmosf{\'e}ricas, Universidad de S{\~a}o Paulo, Rua do Mat{\~a}o, 05508-900, S{\~a}o Paulo-SP, Brazil\\
$^{7}$ Joint Institute for Nuclear Astrophysics - Center for the Evolution of the Elements (JINA-CEE)\\
}

\date{Accepted Received ; in original form }

\begin{document}

\maketitle\label{firstpage}

\begin{abstract}
  
We study the evolution of oxygen abundance radial gradients as a
function of time for the Milky Way Galaxy obtained with our {\sc
Mulchem} chemical evolution model.  We review the recent data of
abundances for different objects observed in our Galactic disc. We
analyse with our models the role of the growth of the stellar
disc, as well as the effect of infall rate and star formation
prescriptions, or the pre-enrichment of the infall gas, on the time
evolution of the oxygen abundance radial distribution. We compute the radial gradient of abundances within the {\sl disk}, and its corresponding evolution, taking into account the disk growth along time. We compare our predictions with the data compilation, showing a
good agreement. Our models predict a very smooth evolution
when the radial gradient is measured within the optical disc with a slight flattening of the gradient from $\sim -0.057$\,dex\,kpc$^{-1}$
at $z=4$ until values around $\sim -0.015$\,dex\,kpc$^{-1}$ at $z=1$ and basically the same gradient until the present, with small differences between models.  Moreover, some models show a steepening at the last times, from $z=1$ until $z=0$ in agreement with data which give a variation of the gradient in a range from $-0.02$ to
$-0.04$\,dex\,kpc$^{-1}$ from $t=10$\,Gyr until now. The gradient
measured as a function of the normalized radius $R/R_{\rm eff}$ is in
good agreement with findings by CALIFA and MUSE, and its
evolution with redshift falls within the error bars of cosmological
simulations.

\end{abstract}


\section{INTRODUCTION}
\label{intro}

Metals are formed inside stars; therefore it is expected that metal
enrichment in the universe should start soon after the formation of
the first massive/short-lived stars, whose explosive demise would
return to the interstellar medium (ISM) newly synthesized chemical elements heavier than primordial
hydrogen and helium. It is thought
that these ejecta, mixed with the surrounding gas, would help its
cooling and precipitate the appearance of a new generation of stars,
giving rise to the cycle of cosmic chemical evolution.

In a more detailed picture, not all the metals are returned to the ISM
with the same timescales: elements like oxygen, sulfur, calcium or
magnesium are synthesized in high-mass stars and returned to the ISM
after only a few million years, while elements like carbon and
nitrogen are mainly made by intermediate-mass stars whose lifetimes
are much longer, of the order of thousands of million years. In
addition, these chemical elements do not return to the ISM in the same
way: the first ones do it through explosive events known as
core-collapse supernovae (supernovae of Type II) which release about
$10^{51} {\rm erg}$ in mechanical energy, while the second ones do it
in a much more quiescent manner through the stellar envelope ejection
in planetary nebulae events, at velocities of the order of $10\,{\rm
km}\,{\rm s}^{-1}$. Finally, an element like iron is mainly produced
in close binary systems after a complex evolution, and ejected in
thermonuclear explosive events (supernovae of Type Ia) on relatively
long timescales. These different ways that the various elements
return to the ISM likely lead to different ways of mixing with the
surrounding gas and, when the explosive pathway is followed, the
possibility will exist that important quantities of metals can leave
the galaxies where they were formed, to enrich the intergalactic
medium.

These processes are imprinted on the distribution of elemental
abundances in galaxies. The absolute quantities of metals which a
galaxy (or a region of a galaxy) possesses, the relative abundances of
the different elements, and their spatial distribution in a given
galaxy are important constraints for the verification of our models
and scenarios of the formation and evolution of galaxies. One very
effective way of handling and exploiting this information is to
confront derived metal abundances at the present time, that is, from
the observation of objects whose lifetime is short as compared to the
lifespan of a galaxy, with chemical evolution models; then use the most
successful models to infer abundances at previous epochs that can be
compared with those corresponding to older galaxy populations. Once
convergence is achieved, the models can be extrapolated back to even
earlier times and predictions can be made which can be contrasted with
planned observations or numerical simulations.

One of the best known features of spiral discs is the existence of a
radial gradient of abundances \citep{hen99}. This gradient was first
observed in the Milky Way Galaxy \citep[MWG, see in][]{sha83}, and
then in other external galaxies as shown in \citet{mcall85,zar94} and
\citet{vzee98}, and it is now well characterized in our local Universe
\citep{san14}. It can be interpreted as due to differences in the star
formation rate or the gas infall rate across the disc, although other
reasons are also possible \citep[see][ for a detailed
review]{gotz92}. In order to understand the role of the involved
processes, numerical chemical evolution models were developed early
\citep{lf83,dt84,mat89,fer94}. Most of them were able to reproduce the
present state of our Galaxy, as radial distributions of star
formation, gas density, surface density stellar profile and the radial
gradient of most common elements (C, N, O, Fe...), however not all of
them predict the same evolution with time. In fact, as \citet{kop94}
explained, the radial abundance gradients may be only modified by
inflows or outflows of gas, and, besides the possible variations of
the initial mass function (IMF) or stellar yields along the
galactocentric radius, the only way to change a radial gradient of
abundances is to have a star formation rate (SFR) or an infall rate
changing with galactocentric radius.

It was demonstrated a long time ago \citep{lf83,mat89} that an
inside-out scenario of formation of discs, as produced by different
infall rate of gas along the disc, is necessary to form a negative
radial gradient of abundances. However, in order to obtain a value of
gradient as observed, it is also necessary to invoke radial flows of gas or
variable star formation efficiencies, which should be higher in the inner
regions than in the outer ones. Using these assumptions, most models
able to reproduce the present state of the solar neighborhood and the
Galactic disc as a whole, follow one of two trends: a) It is initially
flat and it steepens with time; or b) the gradient is steep, and
flattens with time. For example, models from \citet{dt84,chia97}
predict an initially flat radial distribution of abundances which then
steepens with time. Conversely, models by
\citet{fer94,pra95,mol97,por00,hou00} present a steep initial radial
abundance gradient which flattens with time.  The first scenario is
the consequence of the existence of the disc as initial conditions
combined with an infall of primordial gas which dilutes the metal
enrichment of the disc preferentially in the outer regions. In the
second scenario, however, it is the infall of gas which forms the disc
and contributes to the metal budget since it proceeds from the halo
and is therefore somewhat enriched due to earlier star formation.
\citet{gris18} explain that ``the one-infall model with only
inside-out or radial gas inflows predicts a steepening of the gradient
with time, whereas the same case adding a variable star formation
efficiency predicts a flattening of the gradient with
time". Therefore, it seems evident that the radial gradient of
abundances observed in spirals is dependent upon the scenario of
formation and evolution of the disc.

From an observational perspective, the question of appropriate data
sets against which to compare has been debated vigorously. In
\citet{mol97} we analysed the existing data concerning this question
for the MWG with three types of data: 1) planetary nebulae (PNe) of
different masses (that is, different ages) to analyse the oxygen
abundance radial distributions at different times; 2) open and
globular clusters (OC and GC, respectively), to estimate the global
metallicity in objects of different ages; 3) stellar abundances. Each
data set has its own problems for determining the time evolution of
the radial gradient. PNe are useful for estimating the radial gradient
of oxygen, an element not modified by nucleosynthesis in their
progenitor stars \footnote{Although in some new works about
intermediate star nucleosynthesis, these stars may produce some
quantities of oxygen. }. \citet{mac03} analysed the early data around
this question, showing a radial gradient steeper than the one seen in
the present time, although, due to the error bars, the observational
points mixed with the H{\sc ii} region's data in the plot of O/H {\sl
vs.} galactocentric radius $R$. Besides the large errors, the problem
with PNe is that to estimate their galactocentric distances and their
stellar masses (and therefore their ages) is a process with large
uncertainties.

The OC problems arise from the necessary classification of thin or
thick disc or even halo populations, which requires a previous
kinematic study, before their use to determine a radial distribution
in metallicity. Furthermore, the age determination comes from a fit of
their spectra to stellar models which depend on metallicity and age
simultaneously, which added to differences depending on the use of LTE
or NLTE models, may be uncertain. For these reasons, OC age
determination is not an easy task, although the error bars are smaller
than for PNe. The box enclosing the data corresponding to these
objects basically falls in the same region as the young stars in a
plot of [Fe/H] {\sl vs.} $R$.
 
The estimates of age and metallicity for GC seem less problematic
since it is assumed that they form as a single stellar
population.\footnote{This old statement should be revised, since
recent works conclude that there are at least two stellar generations
in these objects, e.g. \citet{bastian18}}.  Ages and metallicities for stars in a GC are
therefore known more precisely than those in OC. Based on these data,
we concluded more than 20 years ago \citep{mol97} that the radial
gradient of abundances in MWG seems steeper in the past and has been
flattening with time until arriving to the present value. However,
these old objects may also be part of the thick disc or the halo
populations. As for the OC, the kinematic information is essential to
determine their group and therefore to know if the corresponding data
may be included in the study of time evolution of the radial
distributions of disc abundances. In \cite{md05}---hereinafter
MD05---we presented 440 chemical evolution models, with 44 different
galaxy masses and 10 possible star formation efficiencies in each
one. From a MWG-like model, the radial gradient of oxygen abundances
is $\sim -0.20$ dex\,kpc$^{-1}$ at $z=2$, which flattens with time
until reaching the present time ($z=0$) value $\sim -0.05$ to $-0.06$
dex\,kpc$^{-1}$. This behavior was in agreement with observations
available at the time.

In recent years, however, new datasets and results from
cosmological simulations have appeared. In the light of these results,
we think that it is time to revise the question of the evolution of
the radial gradient of abundances.  

In the present work we use the {\sc MULCHEM} models, which
are based in those ones from \citet{md05}. They are 1D models in which
we do not include radial gas flows, as it was done in \citet{cavichia14}. There
it was found that the model with radial gas flows induced by the Galactic bar increases the SFR in the 
bar/bulge-disc transition region. A slight difference in elemental abundances is also noted; the model with gas flows predicts a flatter radial distribution in the outer disc but the change is so small that it probably cannot be detected by the current observations. 

Another mechanism recently
included in other models from the literature is (stellar) radial migration,
usually invoked as one of the secular processes that may modify the disc. It is driven by transient spiral patterns
which rearrange the angular momentum \citep[see][ for a detailed
explanation]{min12a,dan18}. In this way, spiral arms may induce the
migration of stars away from their location of birth, resulting in
variations in the stellar composition with galactocentric
radius. The process may explain the existence of the large observed
dispersions in the distribution of stellar metallicities at a given
galactocentric distance, and, as a certain number of stars more
metal-rich than expected will appear in the outer regions, resulting in a
flattening of the radial distribution of stellar abundances.  Thus,
\citet{sanblaz09} or \citet{min12a}, including radial migration in
their cosmological simulations, obtained a stellar metallicity radial
gradient which flattens outside. \citet{ruiz17b} also demonstrate how metallicity gradients for galaxies with different break types change when considering where stars are born and where they have migrated. They find an average change in $\Delta$[M/H] of 0.05~dex/inner disc scale-length comparing galaxies with radial migration and accretion of stars and galaxies with relatively steep metallicity gradient and under the influence of galaxy mergers which drive much of the migration. Following these simulations, it is expected that the migration affects more to the oldest 
stellar populations \citep[see Fig. 22 from][]{sanblaz09}. 

Radial migration is not included in our models. In our scenario,
the Galaxy has a halo or proto-galaxy and a thin disc. It is necessary
to have the kinematic information of stars to follow their movements,
which would require the development of a new code. To include all details of
the complex structure of the Galactic disc (thin and thick discs, bar,
spiral arms, flares) is beyond the scope of this work. We have
undertaken a project to include these 2D structures in our new model which
will be published in the near future. Here we try to understand the
evolution of the MWG in a more generic way in order to apply the same
type of model in other external galaxies. Moreover, the proportion of
discs showing a flattening at the outer regions is only of $\sim
20-30$\% \citep{sanblaz09,ruiz17} and therefore radial migration
is not an essential mechanism to explain the underlying radial
distributions of discs.  It will be necessary, however, to take possible stellar migration into account when we compare the model
results with the data, mainly when old OC or stars are used, and with
object located in the outer regions of disc, when/where the migration
might be more important, since in this case their abundances may not represent the one of the local gas at their formation
time.

We give a summary of the observations related to different age objects in MWG in Section 2.  For our study, we have generated a new suite of chemical evolution models, described in Section 3, where we have revised all the necessary code inputs.  In Section 4, we will analyse the results corresponding to the MWG model, comparing with data compiled in Section 2. In Section 5 we present a discussion about the evolution of the radial gradient and the conclusions are given in Section 6.

\section{OBSERVATIONAL DATA FOR MWG}
\label{data}
We have done a search of data from the literature that we here divide as corresponding to the present time (H{\sc ii} regions and young stars) or to past times (PNe, OC and stars of different ages).

\subsection{Present Day Data}
\label{today}
\subsubsection{Nebular data: H{\sc ii} data.}

We summarize here our analysis of different recent of H{\sc ii} data
sets from several works in the literature. \citet{rudolph06} present
data for H{\sc ii} regions at galactocentric distances of R $\sim$
10-15\,kpc, which have been observed in the far IR emission lines of
[O III] (52\,$\mu$m, 88\,$\mu$m), [N III] (57\,$\mu$m), and [S III]
(19\,$\mu$m) using the Kuiper Airborne Observatory. In total, 168 sets
of observations of 117 H{\sc ii} regions. The new analysis includes an
update of the atomic constants (transition probabilities and collision
cross sections), recalculation of some of the physical conditions in
the H{\sc ii} regions ($n_{e}$ and $T_{e}$), and the use of new
photo-ionization models to determine stellar effective temperatures of
the exciting stars.

\citet{balser11} measure radio recombination line and continuum
emission in 81 Galactic H{\sc ii} regions at Galactocentric radii of 5
to 22\,kpc and also sample the Galactic azimuth range
$330-60^\circ$. Using their highest quality data (72 objects) they
measured the electron temperature and derived the oxygen abundance from
the calibration given in \citet{sha83}:
\begin{equation}
12+\log({\rm O/H})=9.82(\pm 0.02) -1.49(\pm0.11)T_{\rm e}/10^{4},
\end{equation}
thus deriving an O/H Galactocentric radial gradient of -0.0383$\pm$
0.0074\,dex\,kpc$^{-1}$. Combining these data with a similar survey
made with the NRAO 140 Foot telescope, they get a radial gradient of
$-0.0446\pm 0.0049$\,dex\,kpc$^{-1}$ for a larger sample of 133
nebulae.\footnote{Dividing their sample into three Galactic azimuth
regions produced significantly different radial gradients that range
from $-0.03$ to $-0.07$\,dex\,kpc$^{-1}$, depending on the azimuthal
angle. These inhomogeneities suggest that metals are not well mixed at
a given radius. They stress the importance of homogeneous samples to
reduce the confusion of comparing datasets with different
systematics. Standard Galactic chemical evolution models are typically
spatially 1D, computing the chemical evolution along the radial
dimension as a function of time. Although this subject is beyond the
scope of this work, our future models will consider azimuthal
evolution as well \citep[see][ in preparation]{wekesa18}.}

\citet{alba} have done observations with the William Herschel
Telescope in the Los Muchachos Observatory (La Palma) for 9 H{\sc ii}
regions. They joined them to other literature data, and reanalysed all
in a homogeneous manner. In total they obtain abundances for 23
objects, covering the galactocentric radius from 11\,kpc to 18\,kpc
i.e., in the outer disc. They measure the radial gradient for O, N, S,
Ar and also for He and N/O, obtaining the following values:
$-0.0525(\pm 0.009$), $-0.080(\pm 0.019)$, $-0.106(\pm 0.006$),
$-0.074(\pm 0.006$), $-0.0005(\pm 0.0019$) and $-0.041(\pm 0.006$)
\,dex\,kpc$^{-1}$, respectively. Similarly, \citet{est17} analyse
H{\sc ii} regions also for the outer regions of disc using the Gran
Telescopio Canarias (GTC) telescope, finding a radial gradient for O
of $-0.0399$\,dex\,kpc$^{-1}$.

\begin{figure}
\includegraphics[width=0.35\textwidth,angle=-90]{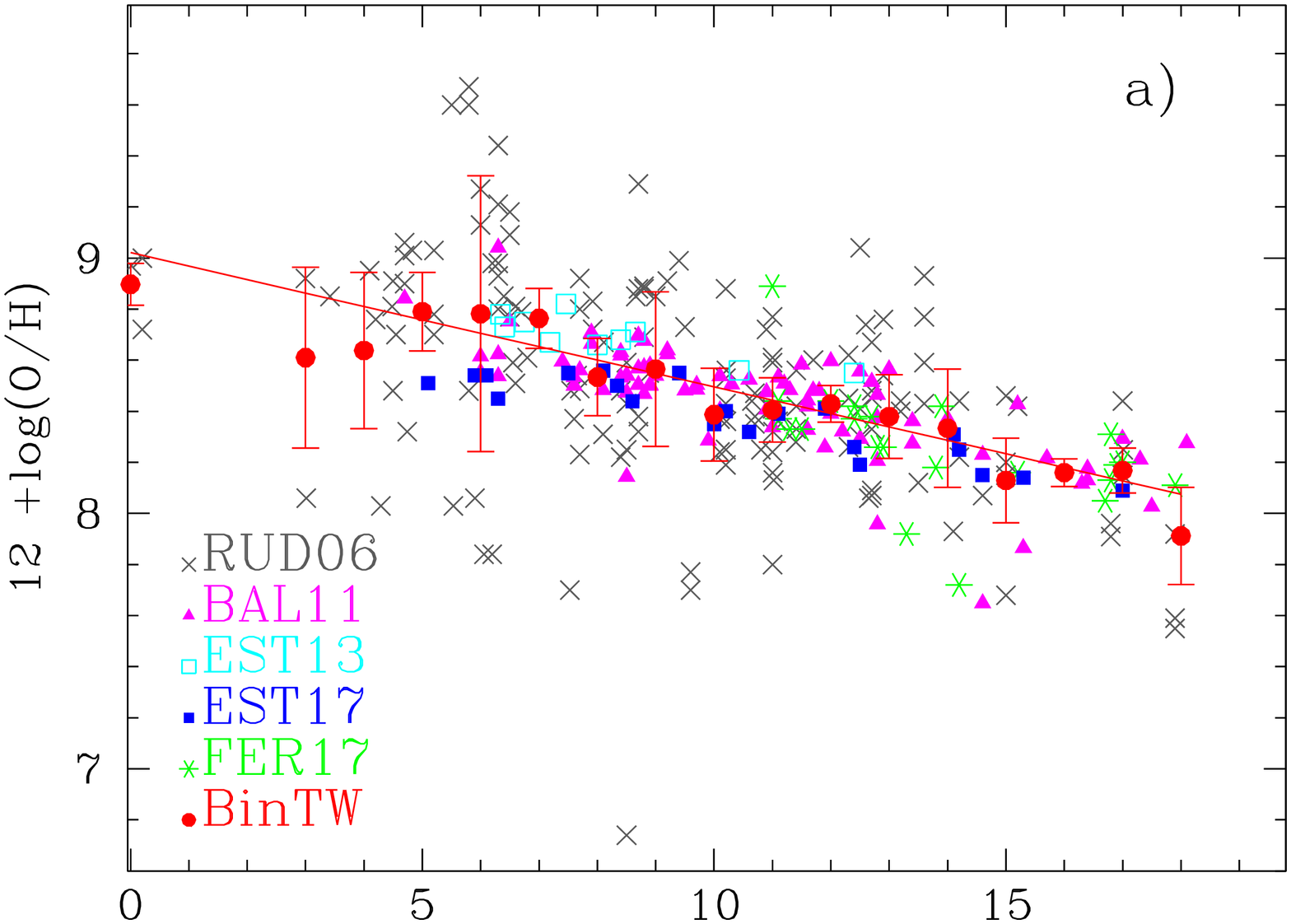}
\includegraphics[width=0.35\textwidth,angle=-90]{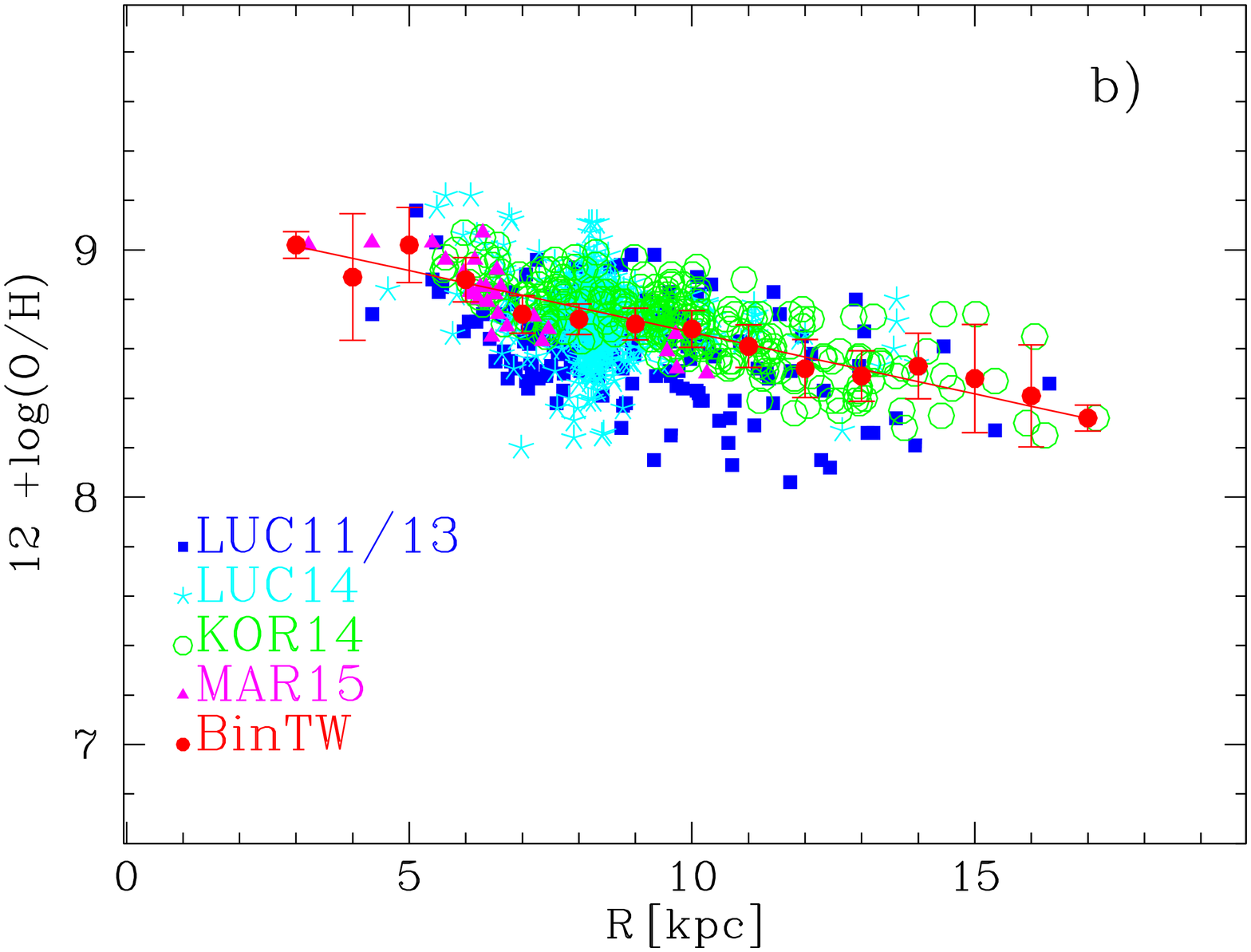}
\caption{The present time radial distribution of the oxygen abundances, $12+\log{(\rm O/H)}$, for MWG with data from different authors: a) for H{\sc ii} regions from \citet{rudolph06,balser11,est17,alba} (RUD06, BAL11, EST17, FER17 respectively) as labelled in the legend; b) for Cepheids stars from \citet{luck11,luck13,luck14,kor14,mar15} (LU11/13, LU14, KOR14 and MAR15 respectively) as labelled in the legend. Our binned results (BinTW) are shown in both panels as red full dots with the least square straight line plotted over the points in each case.}
\label{ohr_obs_hii_cef}
\end{figure}

We have combined all these data in Fig.~\ref{ohr_obs_hii_cef}, panel
a) and performed a binning of all data to have a point for each
kpc. These results are given in Table~\ref{binned}, column 2, with the
corresponding statistical error resulting from the binning
process. We have data from $R=0$ out to 18\,kpc. It is possible that,
in the process of binning, data coming from different samples (obtained in some cases with different techniques) are adding a
systematic error. However, we consider that this systematic error is
already taken into account in the shown dispersion of data and the
corresponding statistical error obtained in the binning of
points. This statement is also valid for the following subsections, where we discuss other objects that are relevant to obtaining oxygen abundances.

\begin{table*}
\caption{Radial distribution of oxygen abundances for different object binned from data of different authors.}
\begin{center}
\begin{tabular}{cccccccccc}
\hline
 & H{\sc ii} & Cepheid & PNe & Young & Intermediate & Old  & Young  & Intermediate  & Old  \\
  & Regions    & Stars  &  &  age OC & age OC & age OC  & age Stars &  age Stars   & age Stars \\
\hline
R & \multicolumn{9}{c}{12+log(O/H)}\\
(kpc) & \multicolumn{9}{c}{(dex)} \\
\hline
0	&	8.90	$\pm$	0.12	&			&			&			&			&			&			&			&			\\
1	&			&			&	8.67	$\pm$	0.12	&			&			&			&			&			&			\\
2	&			&			&	8.72	$\pm$	0.12	&			&			&			&			&			&			\\
3	&	8.61	$\pm$	0.28	&	9.02	$\pm$	0.05	&	8.72	$\pm$	0.10	&			&			&			&			&			&			\\
4	&	8.64	$\pm$	0.23	&	8.89	$\pm$	0.15	&	8.72	$\pm$	0.12	&			&			&			&	8.55	$\pm$	0.05	&	8.55			&	8.55			\\
5	&	8.79	$\pm$	0.11	&	9.02	$\pm$	0.10	&	8.68	$\pm$	0.11	&	8.95	$\pm$	0.07	&			&			&	8.44	$\pm$	0.23	&	8.57	$\pm$	0.10	&	8.40	$\pm$	0.10	\\
6	&	8.78	$\pm$	0.15	&	8.88	$\pm$	0.07	&	8.62	$\pm$	0.09	&	8.81	$\pm$	0.08	&			&			&	8.69	$\pm$	0.09	&	8.70	$\pm$	0.06	&	8.58	$\pm$	0.07	\\
7	&	8.76	$\pm$	0.10	&	8.74	$\pm$	0.06	&	8.70	$\pm$	0.07	&	8.75	$\pm$	0.11	&	8.97	$\pm$	0.05	&			&	8.67	$\pm$	0.06	&	8.68	$\pm$	0.05	&	8.62	$\pm$	0.06	\\
8	&	8.53	$\pm$	0.09	&	8.72	$\pm$	0.06	&	8.64	$\pm$	0.08	&	8.61	$\pm$	0.08	&	8.68	$\pm$	0.10	&	8.80	$\pm$	0.14	&	8.66	$\pm$	0.05	&	8.65	$\pm$	0.05	&	8.62	$\pm$	0.06	\\
9	&	8.62	$\pm$	0.09	&	8.70	$\pm$	0.06	&	8.64	$\pm$	0.13	&	8.70	$\pm$	0.07	&	8.69	$\pm$	0.08	&	8.61	$\pm$	0.11	&	8.62	$\pm$	0.05	&	8.58	$\pm$	0.06	&	8.55	$\pm$	0.07	\\
10	&	8.39	$\pm$	0.10	&	8.68	$\pm$	0.06	&	8.57	$\pm$	0.10	&	8.61	$\pm$	0.09	&	8.80	$\pm$	0.10	&	8.70	$\pm$	0.05	&	8.61	$\pm$	0.07	&	8.63	$\pm$	0.07	&	8.61	$\pm$	0.08	\\
11	&	8.40	$\pm$	0.08	&	8.61	$\pm$	0.07	&	8.47	$\pm$	0.12	&	8.55	$\pm$	0.14	&	8.49	$\pm$	0.11	&	8.56	$\pm$	0.09	&	8.64	$\pm$	0.06	&	8.69	$\pm$	0.08	&	8.63	$\pm$	0.14	\\
12	&	8.43	$\pm$	0.07	&	8.52	$\pm$	0.08	&	8.71	$\pm$	0.12	&	8.40	$\pm$	0.09	&	8.50	$\pm$	0.08	&	8.65	$\pm$	0.13	&	8.57	$\pm$	0.07	&	8.54	$\pm$	0.07	&			\\
13	&	8.38	$\pm$	0.09	&	8.49	$\pm$	0.08	&	8.33	$\pm$	0.12	&	8.60	$\pm$	0.10	&	8.44	$\pm$	0.10	&	8.64	$\pm$	0.05	&	8.56	$\pm$	0.08	&	8.54	$\pm$	0.08	&			\\
14	&	8.33	$\pm$	0.13	&	8.53	$\pm$	0.09	&	8.55	$\pm$	0.13	&	8.60	$\pm$	0.11	&	8.55	$\pm$	0.06	&		&	8.48	$\pm$	0.09	&			&			\\
15	&	8.13	$\pm$	0.11	&	8.48	$\pm$	0.13	&	8.53	$\pm$	0.24	&	8.65	$\pm$	0.06	&			&			&	8.63	$\pm$	0.13	&			&			\\
16	&	8.16	$\pm$	0.07	&	8.41	$\pm$	0.13	&			&			&	8.38	$\pm$	0.05	&	8.53	$\pm$	0.05		&			&			&			\\
17	&	8.17	$\pm$	0.09	&	8.32	$\pm$	0.05	&			&			&		&			&	8.53	$\pm$	0.05	&			&			\\
18	&	7.91	$\pm$	0.16	&			&			&			&	8.47	$\pm$	0.05 	&			&			&			&			\\
19 &  &  &  &  & 8.73	$\pm$	0.05 &  & & & \\ 
\hline
\end{tabular}
\end{center}
\label{binned}
\end{table*}
We give in Table~\ref{grad_OH} the radial gradient of oxygen
abundances given by different authors. Each reference is in column 1,
with the corresponding radial gradient of oxygen abundances in column
2, and the radial range covered in column 3. The resulting radial
gradient obtained after the binning process with all points is also
given (this work) in that table at the end of each block for each
object type.

\begin{figure}
\includegraphics[width=0.35\textwidth,angle=-90]{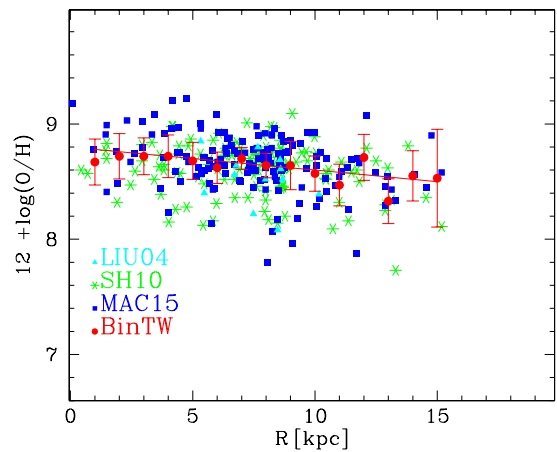}
\caption{The radial distribution of the oxygen abundances, $12+\log{({\rm O/H})}$, for PNe in the MWG. Data are from \citet{liu04,stan10} and \citet{mac15}, denoted by LIU04, SH10 and MAC15, respectively, as labelled in the legend. Our binned results are shown as red full dots with the least square straight line plotted over the points.}
\label{ohr_obs_pn}
\end{figure}

\begin{table}
\caption{Radial gradient of Oxygen abundances from different sets of observational data}
\begin{center}
\begin{tabular}{lcccc}
\hline
Author & Time/age &$\nabla$ [O/H] & radial range & N \\
       & Gyr & (dex\,kpc$^{-1}$) &  (kpc) & \\
\hline
\multicolumn{5}{c}{Present time} \\
\hline
\multicolumn{5}{c}{H{\sc ii} Regions} \\
\hline
RUD06 & Present &$-0.0543 \pm 0.008$ & 3--18 & 151  \\
BAL11 & Present & $-0.0446 \pm 0.0051$  & 4--18 & 81 \\
EST13& Present & $-0.0381 \pm 0.0087$  & 6-13 &10 \\
EST17& Present & $-0.0399 \pm 0.0044$  & 5--17& 20 \\
FER17& Present & $-0.0525 \pm 0.0189 $  & 11-18 & 22\\
This work & Present & $-0.048 \pm 0.005$ & 0--18 & 284 \\
\multicolumn{5}{c}{Cepheids} \\
\hline
LUCK11/13 & Present & $-0.0430 \pm 0.0065$ & 4--17 & 168 \\
LUCK14 & Present &  $-0.0225 \pm  0.0047$ & 4--14 & 815\\
KOR14 & Present & $-0.0529 \pm 0.0083$ & 5-18 & 271\\
MAR15 & Present & $-0.080 \pm 0.010$& 3-11 & 26\\
This work & Present & $-0.0498 \pm 0.05$ & 3--17 & 1280 \\
\hline
\multicolumn{5}{c}{Past time} \\
\hline
\multicolumn{5}{c}{PNe} \\
\hline
LIU04 & 2-4 & $-0.048 \pm 0.048 $ & 5--11 & 12  \\ 
SH10 & all & $-0.024 \pm 0.019 $ & 1--15& 125 \\
& Type I    & $-0.035 \pm 0.024 $ & & \\
& Type II    & $-0.023 \pm 0.005 $ & & \\
&Type III   & $-0.011 \pm 0.013 $ & & \\
MAC15 & 2-4 & $-0.026 \pm 0.006$ & 0--16 & 149 \\
This work & 2-4 & $-0.020 \pm 0.050$  & 0--16 & 286\\
\multicolumn{5}{c}{Open Clusters} \\
\hline
CAR11 $+$ Lit &  & $-0.049\pm 0.059$& 8-14 & 43\\
& $\tau \le 2$& $-0.0210\pm 0.0175$& 7--14& 39\\
& $2< \tau \le 8$ & -0.060 $\pm$ 0.041& 9-11 & 2 \\
& $\tau > 8$  & $+0.005\pm 0.000$& 8--12& 2\\
YONG12 & &$-0.027\pm 0.008$ & 6--21 & 39 \\               
& $\tau \le 2$ & $-0.085\pm 0.026$&8--13& 13\\
& $2< \tau \le 8$ & -0.0204 $\pm$ 0.008 & 6-20 & 24\\ 
& $\tau > 8$& $-0.083\pm 0.000$ &8--11 & 2\\               
FRIN13& & $-0.044 \pm 0.019$ & 7--14 & 28\\
& $\tau \le 2$ & $-0.036 \pm 0.021$& 8-14 & 22\\
& $2< \tau \le 8$ & -0.106 $\pm$ 0.032 & 8-13 & 5 \\
& $\tau > 8$  & & 9 & 1\\
CUN16  & & $-0.031\pm 0.009$ &  6--17 & 28 \\
& $\tau \le 2$& $-0.022\pm 0.017$ & 6--15& 7 \\
& $2< \tau \le 8$ & -0.030 $\pm$ 0.009 & 8-14 & 7 \\
& $\tau > 8$ & $-0.0354\pm 0.045$ & 8--16& 14\\
MAG17 & & $-0.027\pm 0.0085 $ & 5-17 &  10 \\
 & $\tau \le 2$ & $-0.116\pm 0.034 $ & 5--7 & 8  \\  
 & $2< \tau \le 8$ & -0.060 $\pm$ 0.000 & 10-17 & 2 \\ 
 & $\tau > 8$ &  & & 0  \\   
This work  & $\tau \le 2 $ &$-0.030\pm 0.001$ & 4--15 & 89\\
& $2< \tau \le 8$ & -0.027 $\pm$ 0.004 & 6-20 & 42 \\
 & $\tau > 8 $ & $ -0.028\pm 0.016$ & 8--16 & 19\\
\multicolumn{5}{c}{Stars} \\
\hline
EDV93&  $\tau \le 2$ & $-0.019 \pm 0.050$ & 7--9 & 18 \\
 & $2<\tau \le 8 $ & $-0.0381\pm 0.020$ & 6--11 & 99 \\
 & $\tau > 8 $ & $-0.00394 \pm 0.025$ & 4--11 & 44 \\
CASA11 & $\tau \le 2$&  $-0.039\pm 0.004$ & 4--11 &4680 \\
 & $2<\tau \le 8 $& $-0.0419\pm 0.003$ & 4--13& 4203\\
 & $\tau > 8 $& $+0.0076 \pm 0.0056$ & 4--12& 1326\\
BERG14 & $ \tau \le 2$ & $+0.163 \pm 0.122$  & 7--9& 6 \\
 & $2.5 <\tau \le 8 $ & $-0.0642\pm 0.048$ & 7--10& 61 \\ 
 & $\tau > 8 $ & $+0.057 \pm 0.422$ &  7--9 & 8\\
AND17 &  $\tau \le 2$  &  $-0.033 \pm 0.005$  & 5--16 & 151\\
 & $2 <\tau \le 8 $& $-0.0253\pm 0.0064$& 6--12 &  273\\
& $\tau > 8 $ &  $+0.0724 \pm 0.054$& 5--7 &  51\\
This work& $\tau \le 2$ & $-0.040\pm 0.018$ & 4--16& 4851\\
&  $2 <\tau \le 8 $ & $-0.028\pm 0.002$ & 4--14 & 4636\\
& $\tau > 8 $& $ 0.005\pm 0.012$ & 4--12 & 1429\\
\hline
\label{grad_OH}
\end{tabular}
\end{center}
\end{table}

\subsubsection{Stellar data: Cepheids}

The best way to obtain abundances for present time from stars is
measuring the ones of very young luminous stars such as
Cepheids. \citet{luck13} derive oxygen abundances for a large sample
of Cepheids (103) using the near-IR triplets 777.4\,nm and 844.6\,nm
from an NLTE (Non Local Thermodynamic Equilibrium) analysis. The
spectra used in this analysis are the same as those used in
\citet{luck11b} and \citet{luck11} as the southern Cepheid
sample. Distances are taken from these two works. More recently,
\citet{luck14} obtain an O gradient of $-0.042$\,dex\,kpc$^{-1}$ for a
set of young luminous stars, which reduces to
$-0.033$\,dex\,kpc$^{-1}$ for the subsample without Cepheids.

\citet{kor14} also did an NLTE analysis of the IR oxygen triplet for a
large number of Cepheid star spectra. Together with the data from
\citet{luck13} they obtain a gradient of $-0.058$ dex\,kpc$^{-1}$ for
O. In turn, \citet{mar15} have obtained elemental abundances in 27
Cepheids, the great majority situated within a zone of galactocentric
distances ranging from 5 to 7\,kpc. The data, combined with data on
abundances in the very central part of our Galaxy taken from the
literature, show that iron, magnesium, silicon, sulfur, calcium and
titanium LTE (Local Thermodynamic Equilibrium) abundance radial
distributions, as well as the NLTE distribution of oxygen, reveal a
plateau-like structure or even positive abundance gradient in the
region extending from the Galactic Centre to about 5\,kpc.

\citet{lem13} took high-resolution spectra to measure the abundances
of several light (Na, Al), $\alpha$ (Mg, Si, S, Ca), and heavy
elements (Y, Zr, La, Ce, Nd, Eu) in a sample of 65 Milky Way
Cepheids. Combining these results with accurate distances allows us to
determine the abundance gradients in the Milky Way. Their data,
however, do not include O abundances. The same issue occurs with the
new set of homogeneous measurements of Na, Al, and three
$\alpha$-elements (Mg, Si, Ca) for 75 Galactic Cepheids obtained by
\citet{gen15}. These measurements were complemented with Cepheid
abundances provided by the same group or available in the literature,
resulting in a total of 439 Galactic Cepheids. Accurate galactocentric
distances based on near-infrared photometry are also available for all
the Cepheids in the sample. They cover a large section of the Galactic
thin disc ($4.1 \le R \le 18.4$\,kpc). It is found that these five
elements display well-defined linear radial gradients and modest
standard deviations over the entire range of Galactocentric distances.

In a similar way to the previous subsection for H{\sc ii} data, we
have used data from the above cited authors
\citep{luck11,luck13,luck14,kor14,mar15}, joined and represented in
panel b) of Fig.~\ref{ohr_obs_hii_cef}. We have binned all of them
into a single distribution, given in Table~\ref{binned}, column 3. We
have points between 3 and 17\,kpc. As it is shown in
Table~\ref{grad_OH}, the radial gradient of O abundances is similar
for H{\sc ii} regions and Cepheids stars, with a value $\sim
-0.05$\,dex\,kpc$^{-1}$.

\subsection{Past times}
\subsubsection{Planetary Nebulae}

\citet{mac03} did estimates of the time variation of the O/H radial
gradient in a sample containing about 240 nebulae with accurate
abundances located in the Galactic disc. For most of these nebulae they had O
abundances from the samples of \citet{macqui99} and \citet{mackop94},
although about 40 new nebulae were included. In this case, the radial
gradient for PNe associated to the youngest ages (present time) was
$\sim -0.06$\,dex\,kpc$^{-1}$. These results were consistent with a
flattening of the O/H gradient, roughly from $-0.11$ dex\,kpc$^{-1}$ to
$-0.06$ dex\,kpc$^{-1}$ during the last 9\,Gyr, or from
$-0.08$\,dex\,kpc$^{-1}$ to $-0.06$\,dex\,kpc$^{-1}$ during the last
5\,Gyr.

Other PNe data were obtained from \citet{liu04}, who list elemental
abundances for 12 Galactic PNe. Abundance analyses were carried out
using both strong collisionally excited lines (CELs) and weak optical
recombination lines (ORLs) from heavy element ions. Using the
distances from \citet[][ hereinafter SH10]{stan10}, these data give a
radial gradient of $-0.048$ \,dex\,kpc$^{-1}$, in good agreement with
the above data.

However, the most recent findings of PNe data give a smooth evolution
with time such as in \citet{hen10}, with only a slight flattening of
the gradient, or even a constant value or a steepening with time as in
SH10. This last one gives an average radial gradient of $\sim -0.023$
\,dex\,kpc$^{-1}$. By dividing their sample by types, SH10 found O/H
gradients of $-0.035$, $-0.023$ and $-0.011$ \,dex\,kpc$^{-1}$, for
types I, II and III, respectively, of which the first type corresponds
to the youngest thin disc PNe and the last one the oldest bin. This
implies that the gradient steepens slightly with time.

\citet[][ hereinafter M13]{mac13} reached a similar conclusion, that
the O/H radial gradient is basically the same from 5--6\,Gyr ago until
now. These authors worked with a large sample of data, and used three
different techniques to derive the ages of the progenitor stars. This
allowed them to divide the sample in two or more groups according to
the ages. They found an average O/H radial gradient of $\sim
-0.04$\,dex\,kpc$^{-1}$ and conclude that it has not changed
appreciably \footnote{Depending on the method and on the chosen
subsample, a change from $-0.07$ to $-0.03$ \,dex\,kpc$^{-1}$ from the
old to the young bin is possible, but a contrary change is equally
possible.}. Therefore, within the uncertainties, the O/H gradients from
PNe could be not very different from the gradients observed in younger
objects (H{\sc ii} and Cepheids). \citet[][ hereinafter M16]{mag16} obtain similar results for some
nearby spiral galaxies, M\,33, M\,31, M\,81 and NGC\,300, from the
comparison of the radial distributions of oxygen abundances given by
H{\sc ii} regions and PNe, respectively, without finding any evidence
of evolution of the radial gradient with time.

\citet{mac15} have compiled a large number of PNe abundances (265)
giving a radial gradient of $-0.02$ to $-0.05$\,dex\,kpc$^{-1}$. In
this case they divided the sample in different groups of height $|z|$
over the disc, associating those with the highest $|z|$ to the lowest
mass or oldest objects, in comparison with the closest ones, that will
be the youngest. According to this scheme, the radial gradient is
$-0.0218$\,dex\,kpc$^{-1}$ for the youngest object and
$-0.0268$\,dex\,kpc$^{-1}$ for the oldest (dividing into two bins with
$|z|\le 600$ or $|z|> 600$\,pc, respectively). Taking into account
that the difference between these above values is not large (smaller than the error bar of the data), we may conclude
that the radial gradient of abundances has not changed appreciably in
the last 5--8\,Gyr. In this work we have joined the data from \citet{liu04,stan10} and
\cite{mac15}.  

From \citet{bland16} the scale height for the thin
disc is 220--450\,pc. Following \citet{juric08} it is $\sim$ 300\,pc,
while it is $> 700$\,pc for the thick disc. More recently, from
\citet{mcmill17}, $z_{D,thin}=300$\,pc and
$z_{D,thick}=900$\,pc. Since we are interested in the evolution of the
radial gradient in the thin disc and not in the thick disc or the
halo, we assume a limit in the height scale to define the thin
disc component. Thus, we have selected only objects with $|z| \le
600$\,pc, assuming that these are the true members of the thin disc,
and that objects above this height do not pertain to the thin disc,
but instead are members of the thick disc or halo. This limit
seems to be safe enough, even taken into account the flare of the outer
disc. This same assumption is also taken in the subsection 2.2.3 about
stellar data. Curiously, when we constrain our sample to this height
above the disc, the galactocentric distances of the selected PNe are
all within R$<15$\,kpc, that is they are surely located within the
thin disc at the present time.  We show O abundances in
Fig.~\ref{ohr_obs_pn} and our binned abundances (column 4 from
Table~\ref{binned}). We have points in the radial range
between R=0 and 16\,kpc. For our points we have computed the radial
gradient as $\sim -0.020$\,dex\,kpc$^{-1}$. 

Most of the PNe have ages $< 4$\,Gyr \citep{mac13} and therefore the radial migration is not expected to be important when considering PNe. See the simulations of \citet{kub15}, in particular their Fig. 6, showing that for the youngest stars ($< 4$ Gyr old) the effect of radial migration is very small.

\subsubsection{Open clusters}

For OC practically all data, from earlier publications
\citep{friel02,chen03,mag09} to the more recent \citep{frin13,cun16},
arrive at the same conclusion: the radial gradient of stellar
metallicity was steeper in the past than now. However, most of data
referring to OC estimate this gradient with Fe abundances or with a
metallicity index as [M/H]. \citet{friel02,friel10} measured an [Fe/H]
gradient of $-0.080$\,dex\,kpc$^{-1}$ for the older OC and
$-0.02$\,dex\,kpc$^{-1}$ for the younger ones. \citet{chen03} obtained
the same results.

\begin{figure}
\includegraphics[width=0.35\textwidth,angle=-90]{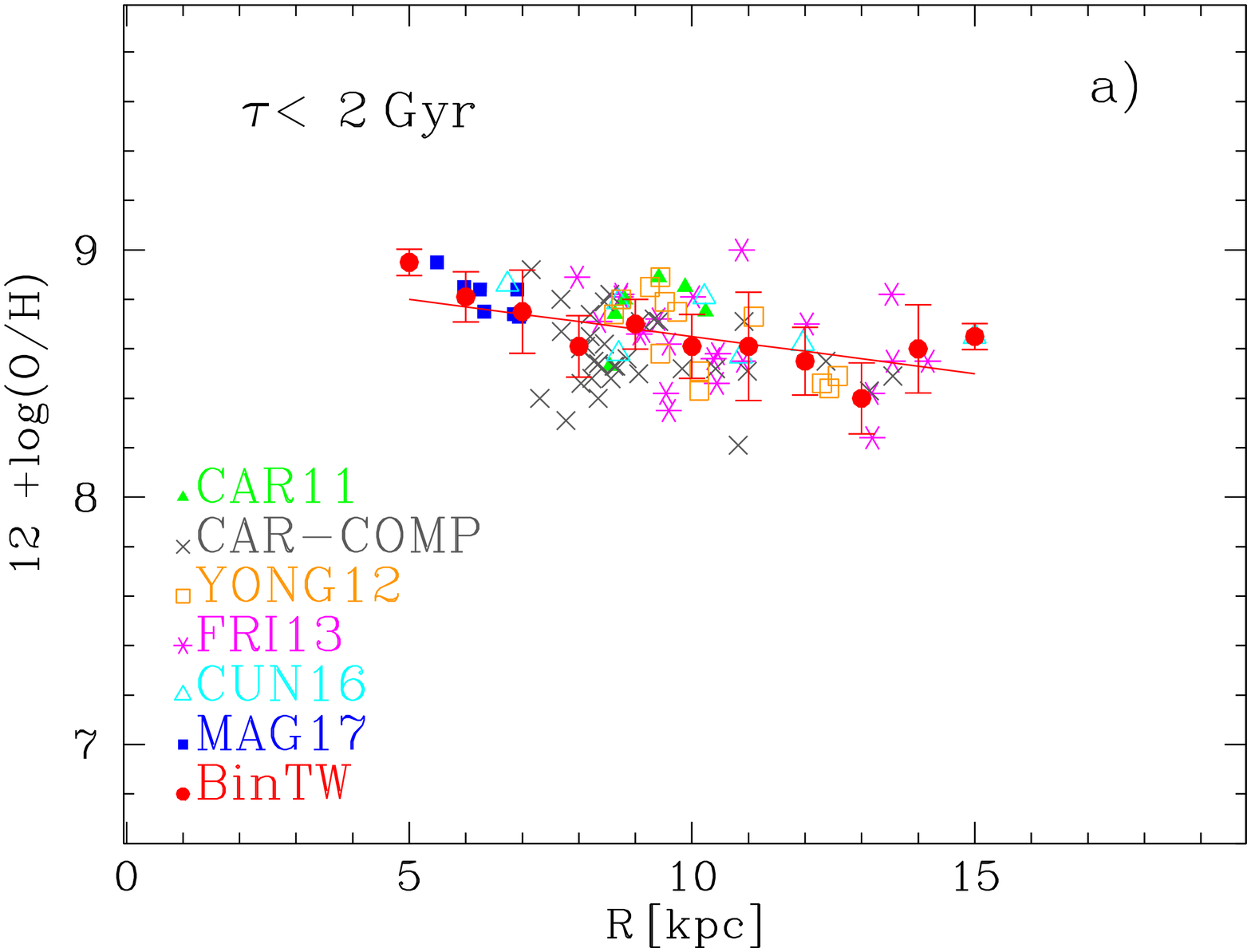}
\includegraphics[width=0.35\textwidth,angle=-90]{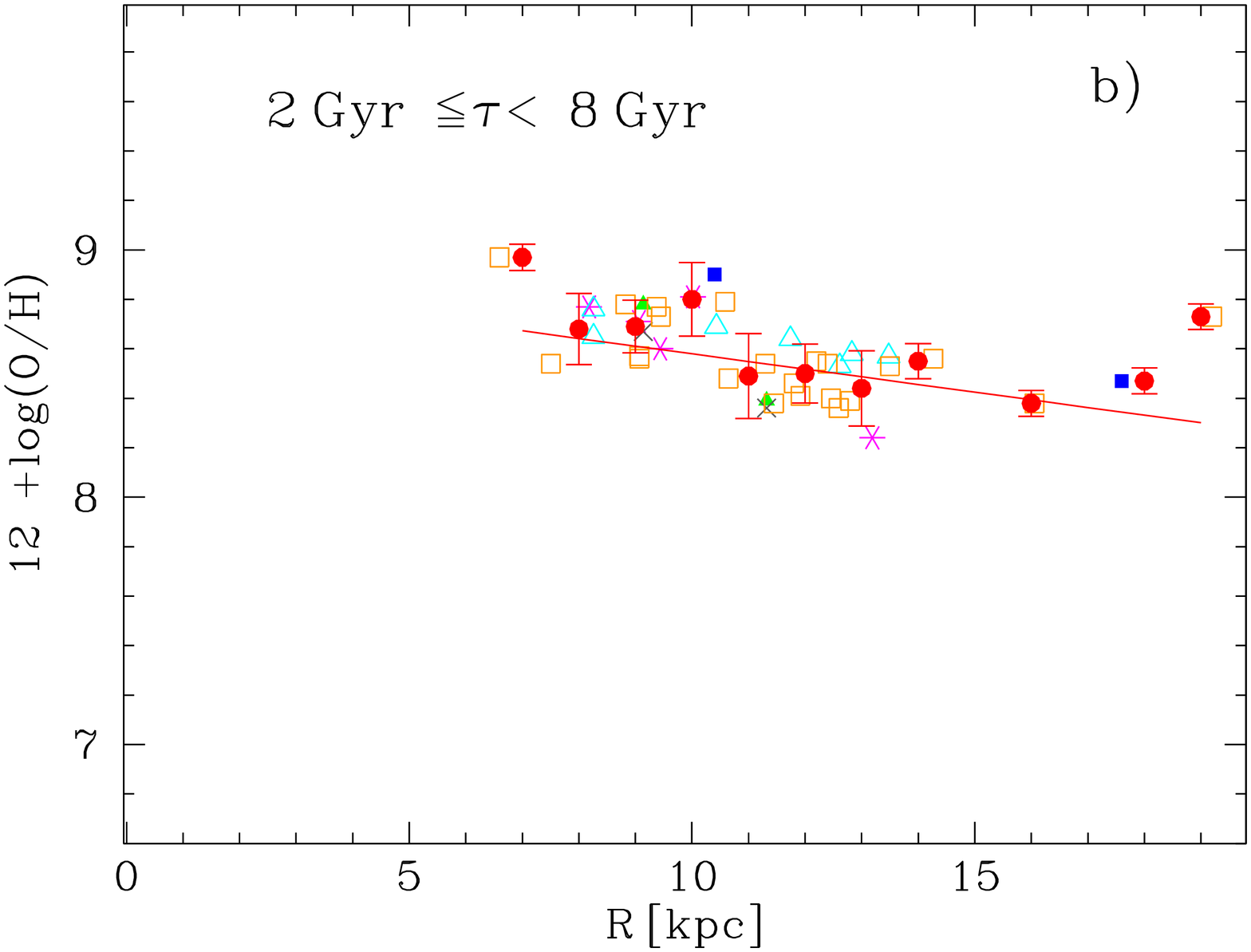}
\includegraphics[width=0.35\textwidth,angle=-90]{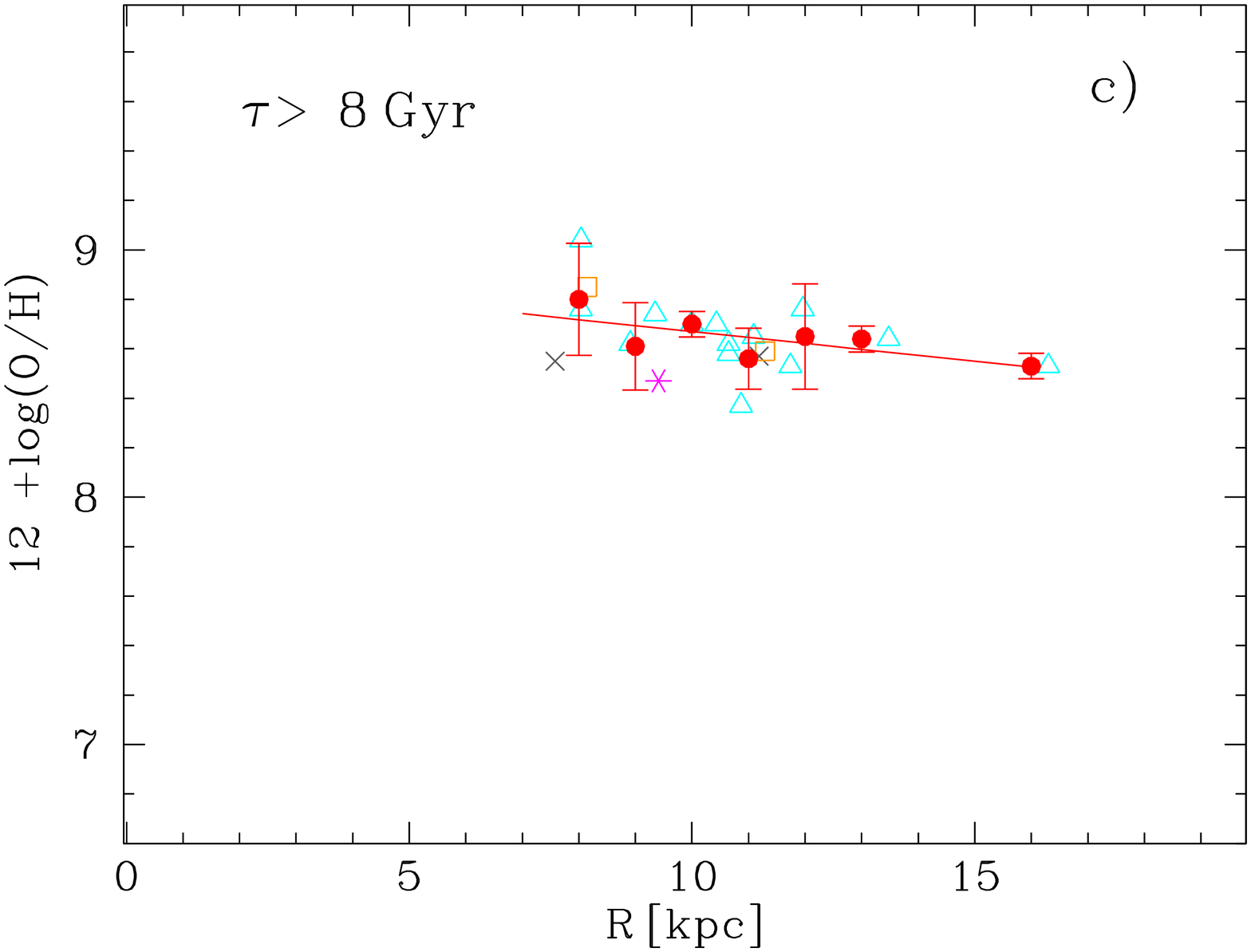}
\caption{The radial distribution of the oxygen abundances, $12+\log{\rm (O/H)}$, for MWG OC of: a) young age; b) intermediate age; and c) old age, with data from different authors: \citet{pan10,car11,yong12,frin13,cun16,mag17}  --as PAN10 and CAR11, FRI13, YONG12, CUN16 and MAG17, respectively-- as labelled in panel a). Our binned results are shown as red full dots in all panels with the least square straight line above the points.}
\label{ohr_obs_oc}
\end{figure}

Similarly,
\citet{pan10,car11,and11,yong12,frin13,red13,cun16,red16,net16,jac16,can16,spi17,mag17}
find that older open clusters use to show a steeper radial gradient
that the youngest ones. This result changes when objects are located
at long galactocentric distances, well outside of the optical radius
i.e., $R> 15$\,kpc. This is specially important when we will confront
the radial gradient of the disc at early times, as predicted by models
(next section), which represent a spiral disc still growing, with data
which maybe form part of the halo or the thick disc. In this case a
flat radial gradient as observed is expected. We give a warning about
this distinction when the study of the radial gradient of spiral discs
is addressed.

Although most of works are devoted to the metallicity, measured as
[Fe/H] or [M/H], some of them estimate the O abundances for OC of
different ages. This way, \citet{pan10,car11} present high-resolution
(R $\sim$ 30 000), high-quality (S/N $\ge$ 60 per pixel) spectra for
twelve stars in four open clusters with the fiber spectrograph FOCES,
at the 2.2 m Calar Alto Telescope in Spain. They employ a classical
equivalent-width analysis to obtain accurate abundances of sixteen
elements: Al, Ba, Ca, Co, Cr, Fe, La, Mg, Na, Nd, Ni, Sc, Si, Ti, V,
and Y. They also derived oxygen abundances by means of spectral
synthesis of the 6300\,\AA\ forbidden line.  They use a compilation of
literature data to study Galactic trends of [Fe/H] and [$\alpha$/Fe]
with galactocentric radius, age, and height above the Galactic
plane. They find no significant trends, but some indicate a flattening
of [Fe/H] at large galactocentric radii, (and also for younger ages in
the inner disc). They also detect a possible decrease in [Fe/H] with
$|z|$ in the outer disc, and a weak increase in [$\alpha$/Fe] with
galactocentric radius.

\citet{yong12} have measured chemical abundances for nine stars in the
old, distant open clusters Be18, Be21, Be22, Be32, and PWM4. Combining
these data with literature, they do a compilation of chemical
abundance measurements in 49 clusters. They confirm that the
metallicity gradient in the outer disc is flatter than the gradient in
the vicinity of the solar neighborhood. They also find that OCs in the
outer disc are metal-poor, with enhancements in the ratios
[$\alpha$/Fe] and perhaps [Eu/Fe]. All elements show negligible or
small trends between [X/Fe] and distance ($\le
0.02$\,dex\,kpc$^{-1}$), except for some elements for which there is a
hint of a difference between the local (R $<$ 13\,kpc) and distant
(R$>$ 13\,kpc) samples, which may have different trends with
distance. There is no evidence for significant abundance trends versus
age (with an age gradient $\le 0.04$\,dex\,Gyr$^{-1}$). They give the
O abundances for 39 of these OC, which we use here.

\citet{frin13} also measured [Fe/H] for globular clusters of different
ages and they also give abundances for $\alpha$-elements as
[$\alpha$/Fe]; although not exactly the same as [O/Fe], we use it to
estimate [O/H]. \citet{cun16} obtained [Fe/H] and [O/Fe] for their
sample of OC, too. Although they give the radial gradients for
different ages only for [Fe/H], we may obtain the gradients for [O/H]
in a similar way.

\citet{mag17} observed open clusters with ages $\tau >$ 0.1\,Gyr. The
sample includes several new clusters: NGC~2243, Berkeley~25, NGC~6005,
NGC~6633, NGC~6802, NGC~2516, Pismis~18 and Trumpler~23 and four
clusters already processed in previous data releases and discussed in
previous papers: Berkeley~81, NGC~4815, Trumpler~20, and
NGC~6705. Most clusters are younger than 2\,Gyr. Only the two outer
ones are older than this.

We summarize all data in Fig.~\ref{ohr_obs_oc}. As we have for the PNe, we have selected objects located at $|z| < 600$\,pc, without any other restriction, and have divided the sample into young, intermediate and old
objects, with ages lower than 2\,Gyr, between 2 and 8 \,Gyr, and older
than 8\,Gyr, respectively, as labelled in each panel. For each
subsample we have binned the abundances, shown as red dots (as in our
previous Fig.~\ref{ohr_obs_hii_cef} and \ref{ohr_obs_pn}), which are
in Table~\ref{binned}, columns 5, 6, and 7. The corresponding radial
gradients for OC of these authors are in Table~\ref{grad_OH}, too. In
the first line for each author we give the gradient as given by them,
or, the one produced using all objects of their samples. In next lines
we give the radial gradients as obtained for us for different age
bins, as indicated.

\begin{figure}
\centering
\includegraphics[width=0.35\textwidth,angle=-90]{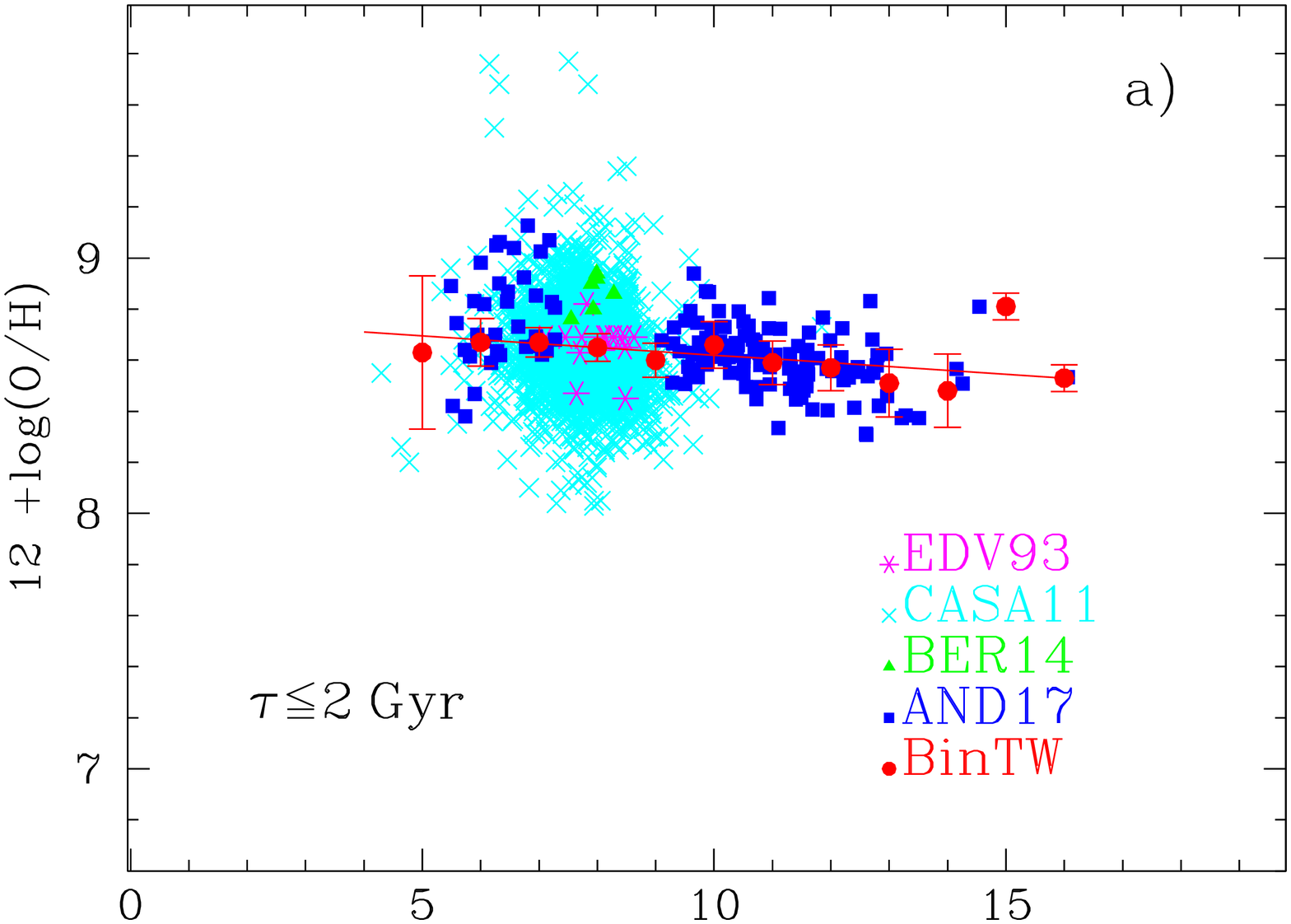}
\includegraphics[width=0.35\textwidth,angle=-90]{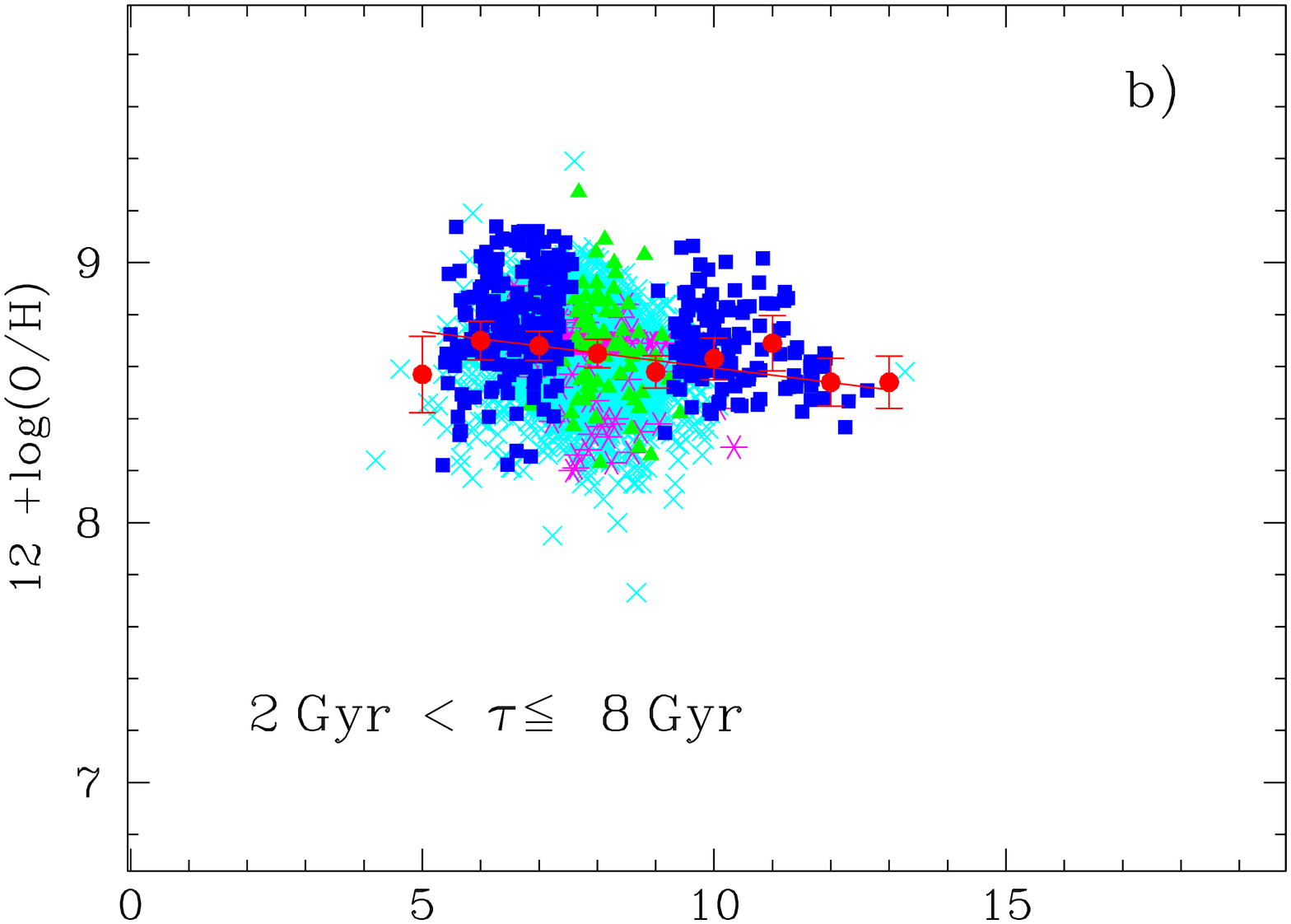}
\includegraphics[width=0.35\textwidth,angle=-90]{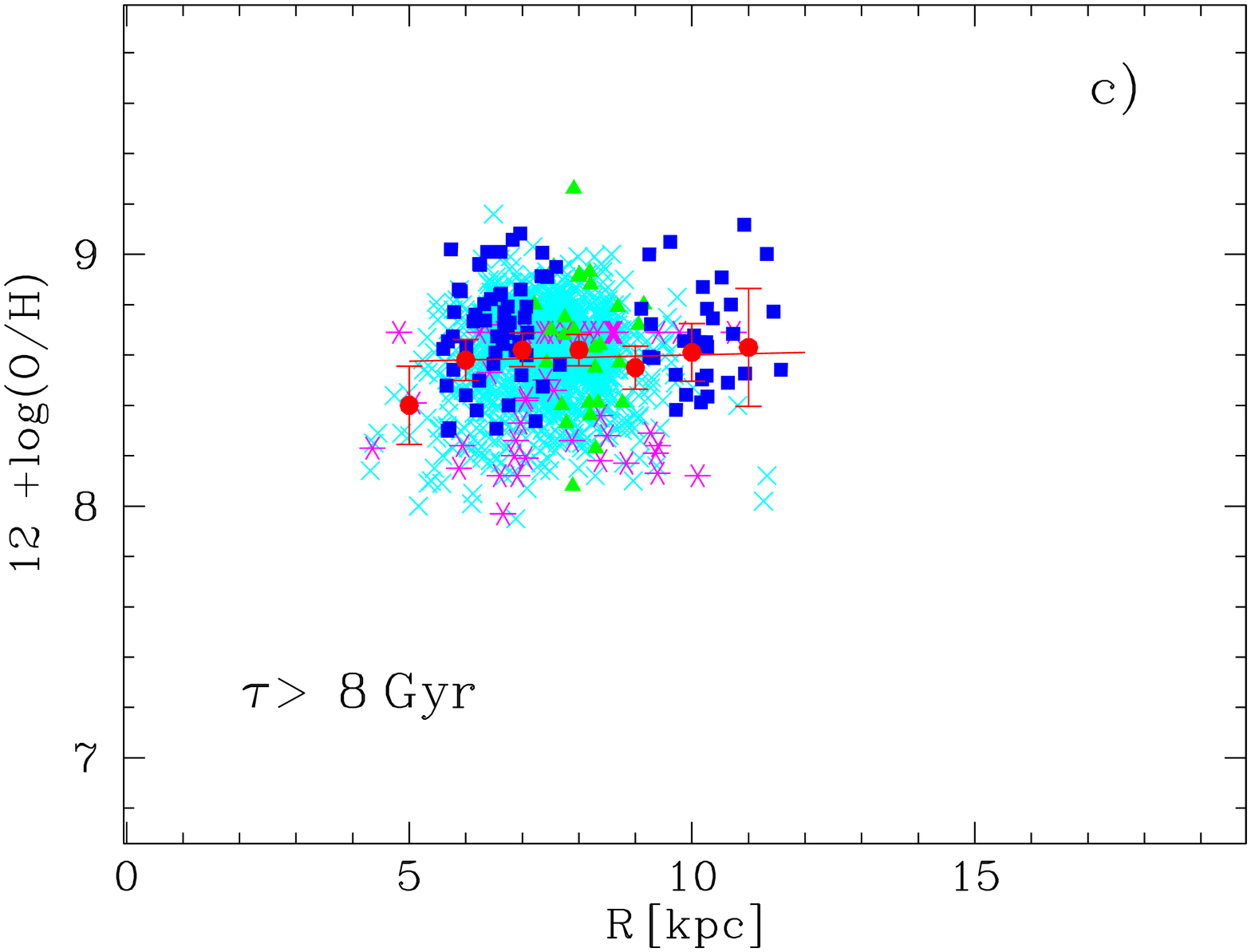}
\caption{The radial distribution of the O abundances, $12+\log{({\rm O/H})}$, for MWG stars of different ages: a) young ($\tau \leq 2$ Gyr); b) intermediate ($2<\tau \leq 8$ Gyr) and c) old stars ($\tau > 8$ Gyr), as indicated in each panel, with data from \citet{edv93,casa11,berg14,anders17}, labelled in panel a) as EDV93,CASA11,BER14 and AND17. Our binned results are shown as red full dots in all panels with the least square straight line plotted over the points.}
\label{ohr_obs_stars}
\end{figure}

\subsubsection{Stellar data}

Finally, we have the stellar data, such as those from
\citet{edv93,nord04,casa11,berg14,xiang15}. Following most of these
authors, stellar abundance data show a flattening of the radial
gradient with time \footnote{For \citet{casa11} and following the
authors discussion for their Fig.18, where they obtain the gradient, we use
the mean orbital radius of stars.}. However, for the oldest age bin
($\sim$ 12-13\,Gyr old) the radial gradient is usually zero, meaning
that the radial gradient started flat, became steeper, and later
flattened again.

\citet{anders17}, from the data for red giants observed by CoRoT and
{\sc APOGEE}, obtain a radial gradient steepening continuously with
time, in disagreement with the other authors. Looking at all these
data in detail, we find some problems for the precise determination of
the radial gradient for the old age bin, since the radial range in
which it is measured is extremely short, preventing the reliable
measurement of a gradient. \citet{berg14} and \citet{xiang15}, who
give the metallicity radial distribution as Fe abundances, divided
their sample into different stellar ages. The radial range for the
whole sample is $\sim 3-4$\,kpc (actually not very wide) which reduces
to $\sim 1-1.5$\,kpc for the oldest stars, i.e., all old objects are
located in a reduced region around the Solar vicinity. It is
impossible to estimate a gradient with such a small radial range.

Similarly, in \citet{anders17} the total sample has a good radial
range (4--14\,kpc), but the oldest bin has a much shorter range at
8--12\,kpc; being in fact a cloud of points around $\sim 9$\,kpc with
a high dispersion. To all these problems we add a cautionary note: the
measurement of the stellar metallicity radial gradient (Z might be
associated to [Fe/H]) might not be the same that the oxygen one, since
they give information from different elements and each element shows a
different evolution. \cite{anders17} summarized these problems to
analyse the evolution of the radial gradient of metallicity as a
function of time by plotting the results of observations for MWG
obtained from PNe, OC, and stars in their Fig.~5.

We here use the data from \citet{edv93,casa11,berg14} and
\citet{anders17} combined into a single set. \citet{edv93} and
\citet{anders17} estimated [O/H] for their stars, while
\citet{casa11} and \citet{berg14} give the iron abundance as
[Fe/H]. However, \citet{berg14} measured [Mg/Fe]. We have then
assumed: [O/Fe]~$\sim$~[Mg/Fe], and then:
\begin{equation}
12+\log({\rm O/H})=[{\rm Mg/Fe}]+[{\rm Fe/H}]+8.69,
\end{equation}
where 8.69 is the solar oxygen abundance \citep{asp09}. Similarly, \citet{casa11} do not provide the oxygen abundances, but they give instead relative abundances of $\alpha$-elements. Assuming $[{\rm O/H}]\sim[\alpha{\rm /Fe}]$ we do:
\begin{equation}
12+\log({\rm O/H})=[\alpha{\rm /Fe}]+[{\rm Fe/H}]+8.69.
\end{equation}
We know that this way of computation is not absolutely precise, but we have checked this assumption for our models and, moreover, this way we may increase the number of points in our samples.

We have divided, as with stars, the sample in three bins of age: 1)
young, with $\tau \le 2$\,Gyr; 2) intermediate, with $2 < \tau \le
8$\,Gyr; and 3) old, with $\tau > 8$\,Gyr. The first subsample would
be compared with the present time data and young OC, the second one
with PNe and intermediate OC, and the oldest one with old OC.
Again we have selected stars with a height $|z|< 600$\,pc. With all
these points, we have obtained the results shown in
Fig.~\ref{ohr_obs_stars} with three panels representing: a) young, b)
intermediate and c) old age stars, respectively. The binned abundances
for each subsample are given (columns 8 to 10) in Table~\ref{binned},
too. The resulting radial gradients, obtained from the binned points
for radial regions 1\,kpc wide as in previous subsections, are shown
in Table~\ref{grad_OH}.

\section{MULCHEM CHEMICAL EVOLUTION MODELS APPLIED TO MWG}
\label{model}
\subsection{Basic description \label{sec:model_descrition}}

Our new {\sc MulChem} chemical evolution models will be described in
detail in Moll{\'a} et al. (in preparation), with details of some of
the used inputs in \citet[][ hereinafter MOL15, MOL16 and MOL17,
respectively]{mol15,mol16,mol17}. Here we summarize the models as
applied to MWG.

In a chemical evolution model, a scenario is assumed in which there is
a given mass of gas in a certain geometric region. This mass is
converted to stars by following an assumed star formation law. A mass
ejection rate appears as a consequence of the death of stars. Often,
some hypothesis concerning gas infall and outflows are included. The
ejected mass depends, therefore, on the remnant of each stellar mass,
on the mean-lifetimes of stars and on the IMF employed. The suite of
models presented here, based on \citet{fer94}, are an update of those
from MD05. We consider a proto-halo with a given initial mass which
fall over the equatorial plane forming the disc.

We start with a radial mass distribution in a proto-halo which is
calculated from the rotation curves given by \citet{sal07} defined in
terms of the total dynamical mass or virial mass, $\log M/{\rm
M}_{\sun}$. Our models are within a range of virial masses: $[5
\times10^{10}-10^{13}]\,{\rm M}_{\sun}$, which implies maximum
rotation velocities in the range $[42-320]$ km\,s$^{-1}$ and would
leads to discs, by the present day, of total mass in the range
$[1.25\times10^{8}-5.3\times 10^{11}]{\rm M}_{\sun}$. The radial
distributions we obtain for various virial masses are shown in Fig.~1
from MOL16. The model to simulate a MWG-like galaxy corresponds to
$\log{M/{\rm M}_{\sun}}=12.01$, which results in a stellar mass of
$M_{\star}\sim 7\,\times 10^{10}$\,\Msun\ by the end of its evolution.

The predicted radial distributions of elemental abundances depend
mainly on three ingredients which are: 1) the infall rate of gas over
the disc; 2) the stellar yields (with the corresponding mean-lifetimes
of stars) and the IMF; and 3) the star formation law. Each one of
these parameters and its role in the radial gradient for elemental
abundances will be shown here. We summarize in the following
subsections these inputs of the model.

In this model we have not included the stellar bar in the centre
nor the spiral arms, that is, we will have an axi-symmetric disc in
which there are no inflows of gas due to the bar as we did in
\citet{cavichia14}. In that work we showed that the radial
distributions are modified when the gas inflows are considered: the
elemental abundances increase in the inner disc, as the star formation
rate does, while the outer disc shows a slightly flatter radial
gradient of abundances. The total radial gradient, taking into account
both effects, is, however, similar to the one found in a model without
the bar. Therefore, we consider that including the bar is important
when the goal is studying structures and details of different parts
within the disc, as the region located near the bulge, where the
effect is more important that further out. The presence of the spiral arms
may also enhance the formation of molecular clouds and subsequent
star formation, and in this way produce arm-interarm or azimuthal
variation in gas elemental abundances. Azimuthal variations
are not easily observed showing differences as smaller as 0.02\,dex
\citep{bovy12}, although recently \citet{sm16b}, studying the spiral
arms in NGC~6754 with VLT/MUSE data, estimate variations as large as
$\sim 0.06$\,dex.  Furthermore, some differences may be also deduced
by statistical methods \citep{sm17}. On the other side, azimuthal
differences could be only visible during the bar phase, tending to
disappear with time \citep{dimat13}. Therefore, an improvement in the instruments is needed, with which we might to detect these
small dispersions in the gas abundances. The effects of the spiral arms are beyond the scope of this work and we will analyse it in a forthcoming work (Wekesa et al. 2018, submitted).

The spiral pattern may also produce radial migration of gas and
stars. As stated in the Introduction, the migration of stars and other
consequence of the existence of a bar and spiral arms, is at present
one of the mechanism claimed to explain the flattening of the radial
gradients of abundances. However, some simulations find only small changes in the radial gradient due to radial migration. For example, \citet{grand15} studied the effects of radial
migration of stars on the [Fe/H] radial gradient computing simulations
which show a clear scattering of the [Fe/H] abundances at all radii,
while the slope of the radial metallicity gradient does not appear to
change, at least in the timescale analysed (1\,Gyr). This effect has
been also reported in other previous models
\citep{sel02,scho09,grand14}.
In any case, it seems that stellar data for young stars may be used
without problem, since in this case radial migration is not
important. It can be seen in \citet{sanblaz09,kub15} that for the
youngest stars ($< 4$\,Gyr old) the effects of the radial migration are
small. Moreover, our objective is analyzing the O abundances. Since
oxygen is mainly produced by massive stars ($M_{*} > 10$\,\Msun) and
because they are short-lived (lifetime shorter than 20\,Myr), such
stars have no time to migrate away from their birth places. Therefore,
the radial O profile from young and intermediate age stars is not
affected by radial migration.  However, if we are going to compare
models with old stars that may have ages of more than 8\,Gyr, then radial
migration may well be important. It would be necessary to compute
hydrodynamical models to simulate how stars migrate, something out of
the scope of the models used here. Therefore, we do not include migration
in this suite of models, but we will take into account this
possibility in the analysis of the results and comparison with
intermediate and old age objects.

\subsection{Galaxy formation and infall rate}

In our scenario, the gas initially in the proto-halo falls to the
equatorial plane where the disc forms. Once we have obtained the
radial distributions of mass in the proto-halo and disc in each
geometrical region at given galactocentric distances, it is possible
to calculate the collapse time in each radius $R$ as:
\begin{equation}
\tau(R)=-\frac{13.2}{\ln{\left(1-\frac{\Delta M_{\textrm{\scriptsize D}}(R)}{\Delta M_{\textrm{\scriptsize tot}}(R)}\right)}}\,[\mbox{Gyr}]
\end{equation}

This collapse timescale for MWG shows now a smoother dependence on
radius and on time than the one used in MD05, such as we show in Fig.2
from \citet{mol16}. The resulting infall rates produced by this
collapse timescale are also different than before: they show a smooth
evolution for disc regions and are stronger for the bulge (see Fig.3
from MOL16). The infall rates for different regions within the disc
show variations only in the absolute values, with a very similar
behavior for all radii. The infall rate is very low in the outer
regions of disc, and stays low with redshift. This implies that the
SFR must be also low for all times. A sharp break in the disc appears,
associated with the dramatic decline in star formation at a given
radius.

\subsection{Stellar Yields} 

The stellar masses are divided into two ranges corresponding to low-
and intermediate-mass stars ($m < 8$\,\Msun), and massive stars ($m >
8$\,\Msun). The first ones eject mainly He$^{4}$, C$^{12,13}$ and
N$^{14,15}$; a small amount of O can be generated in some yield
prescriptions, as well as various s-process isotopes. Massive stars,
in turn, produce C, O, and all the so-called $\alpha$-elements, up to
Fe. The literature of stellar yield generation is a rich one, with
various works differing from one another due to the intrinsically
different input physics to the underlying stellar models. The
supernova-Ia ejecta are taken from the classical model W7 from
\cite{iwa99}, using the supernova rates as given by \citet{rlp00} and
assuming a binary stars ratio of $\alpha=0.20$.

It is necessary to weight these stellar yields by the Initial Mass
Function (IMF). We have addressed this question in MOL15. There we
have computed the same basic chemical evolution model for MWG by using
6 different stellar yields for massive stars, 4 different yields sets
for low- and intermediate-mass stars, and 6 different IMFs. Analyzing
these 144 permutations and comparing their results with the
observational data for our Galaxy disc, we determined which
combinations are most valid in reproducing the data. From these
results, the different IMF $+$ stellar yields combinations produce
basically the same radial distributions for gas (diffuse and
molecular), and stellar $+$ star formation surface densities. The
corresponding radial distributions for the elemental abundances of C,
N and O are very different in their absolute values, with some of them
far from the observational data, while others lie closer. They do,
however, show a similar slope for these radial distributions,
suggesting that the selection of one or other combination is not
particularly useful in modifying the radial abundance gradient. In any
case, there are only 8 combinations of IMF $+$ stellar yields able to
reproduce the MWG data with a high probability ($P> 97$\%). Here we
use the yield combination named GAV-LIM-KRO, with yields given by
\citet{gav05,gav06} and \citet{lim03,chi04}, with the IMF of
\citet{kro02}, which provides a good match to the data. See more
details in MOL15.

\subsection{Prescriptions for the H{\sc i} to H$_{2}$ conversion process}

In {\sc MulChem} the star formation in the disc occurs in two steps:
first, molecular gas forms, and then stars are created by cloud-cloud
collisions or interactions of massive stars with the surrounding
molecular clouds. The formation of both molecular clouds and stars was
treated through the use of efficiencies, considered as free parameters
in MD05. Recently, we have shown in MOL17 that the prescriptions given
in the literature for the formation of molecular clouds may be
adequate to be included in our code. We have checked some of these
possibilities in MOL17, comparing the results obtained for a Galactic
chemical evolution model regarding the evolution of the Solar region,
the radial structure of the Galactic disc, and the ratio between the
diffuse and molecular components, H{\sc i}/H$_{2}$ with the existing
data: the six tested prescriptions successfully reproduce most of the
observed trends. The model proposed by \citet{asc17}, where the
conversion of diffuse gas into molecular clouds depends on the local
stellar and gas densities as well as on the gas metallicity, is the
best model for reproducing the observed data. Therefore we use in this
case the prescription named ASC in MOL17, to create molecular clouds
from the diffuse gas.

\subsection{Computed Models}
\begin{table}
\caption{Description of computed models}
\begin{tabular}{ccccc}
\hline
Name & Infall type & $H_{2}$ prescription & $\epsilon_{h}$ & Reference\\
\hline
MOD1 & MOL16 & ASC & 0.03 &  MOL17\\
MOD2 & MOL16 & STD & 0.03 & MOL17\\
MOD3 & MD05 & STD & 0.01 & MD05\\
MOD4 &  MOL16 & ASC & 0.00  & This work\\
MOD5 &  MOL16 & ASC & 0.10 & This work\\
\hline
\hline
\label{models}
\end{tabular}
\end{table}
The MWG model corresponds to $\log{M_{\textrm{vir}}}=12.01$ and an
efficiency $\epsilon_{s}$ corresponding to NT=4, with
$\epsilon_{s}=\exp^{(-NT^{2}/8)}$. This model was calibrated against
the MWG data in MOL15 and MOL17.

We now compare some different models applied to MWG in which we have
modified some parameters to seek differences over the radial
distribution of O abundance. The suite of models are described in
Table~\ref{models}, in which we have the name of each model in column
1. Column 2 lists the infall of gas prescription used: the one used in
{\sc MulChem} proposed by MOL16 and also used in MOL17, compared with
the old one from MD05. Column 3 gives the type of prescription used to
convert the diffuse gas into molecular gas; we have used only two
possibilities, named: ASC and STD, as defined in MOL17 (see this work
for more details). The STD prescription refers to the classical
efficiency, used in MD05 as a free parameter. Finally, in column 4 we
show the efficiency $\epsilon_{h}$ of the star formation law in the
halo.

This way, comparing MOD1 and MOD2 we may see differences in results
coming from using different prescriptions for the creation of
molecular clouds; comparing MOD2 and MOD3 we see the effect of the
infall rate assumed in our new models compared with the old one from
MD05; comparing MOD1, MOD4 and MOD5, we see results when the
efficiency of star formation in the halo, $\epsilon_{h}$ is changed.

\section{RESULTS}
\label{mwg}

\subsection{The radial range of the abundance radial gradient}
\begin{figure}
\includegraphics[width=0.35\textwidth,angle=-90]{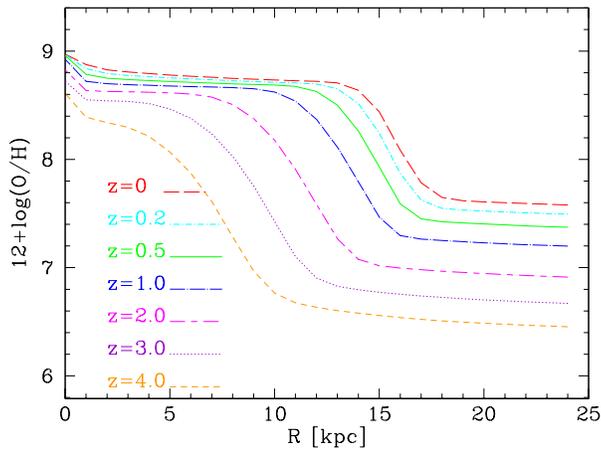}
\caption{The radial distribution of the O abundances, $12+\log{({\rm O/H})}$, for different redshifts, as labelled. The results are for MOD1 and all radial regions.}
\label{oh_z}
\end{figure}

\begin{figure}
\includegraphics[width=0.35\textwidth,angle=-90]{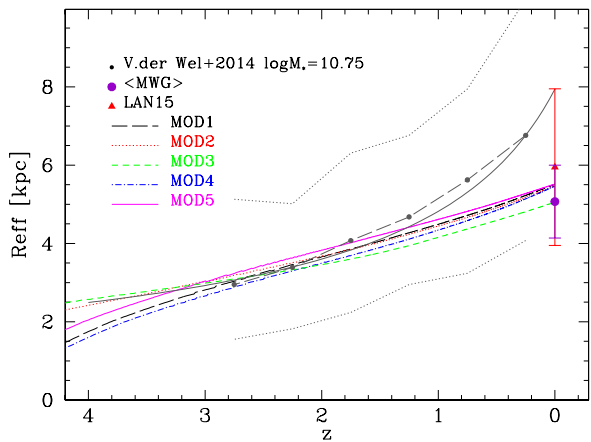}
\caption{Evolution of the effective radius $R_{\rm eff}$ of models with the redshift $z$ compared with data from \citet{vdWel14} for galaxies with stellar mass $\log{M_{\star}}\sim 10.75$, shown as dashed lines with dots (the dotted lines are the curves $\pm \sigma$). The full purple dot represents the estimated effective radius for the MWG at the present time, while the red triangle is the averaged value measured for galaxies of the local Universe with GAMA data \citep{lange15}. Each model is represented with a different type of line and color as labelled.}
\label{reff_z}
\end{figure}
In Fig.~\ref{oh_z}, the oxygen abundance, as $12+\log{({\rm O/H})}$,
for MOD1 is represented as a function of the radius $R$ in kpc, for
seven values of redshift: z= 0, 0.2, 0.5, 1.0, 2.0, 3.0 and 4.0. We
have represented all computed radial regions. For higher redshifts the
radial distributions are apparently very different than now, with a
strong variation with redshift/time. However, we see that there is not
a straight line with only one slope, but there is a clear line until a
radius $R\sim 13-14$\,kpc in $z=0$. Beyond this region there exists a
strong downturn which would correspond to the edge of the stellar
disc. This shape is similar to the one obtained by \citet[][ see
their Fig.5]{wang18} for the density of the Galactic disc. These
authors study the radial density profile of the disc, confirming that
it extends to 19\,kpc, and separating the contributions of thin and
thick disc for each radial bin. They find two breaks at R=11\,kpc and
R=14\,kpc, thus dividing the radial distribution into three segments,
one for $R<11$\,kpc, other for $R$ in the range 16--19\,kpc, and a
middle transition zone between both. The scale lengths are 2.12\,kpc
in the inner region, 2.72\,kpc in the outside, and 1.18\,kpc in the
intermediate one. The outer disc (starting at 14\,kpc) is where the thick
disc component becomes prominent. The thin disc, the dominant
contributor to the inner region, begins to weaken at the first break,
disappearing gradually within the transition region. These authors suggest that the stellar populations of the outer disc
correspond to the thick disc and that the thin disc ends at a shorter
radius. Our breaks, for the present time, occur at radii slightly
higher, the first one at $\sim 14$\,kpc, and the second at $\sim 17$\,kpc. Considering the findings of \citet{wang18}, the radial gradient for the disc must be measured with data from regions located within the first break, since the abundances of outer regions located at $R> 16-17$\,kpc would correspond, in our models, to the thick disc/halo regions, for which a flatter radial gradient is expected.

Apparently, the curves for other redshifts have the same behavior,
with a smooth variation until a given radius and a decreasing after
that. We should note that a variation of size with redshift is
expected in the inside-out scenario of growth of discs. We need,
therefore, to define a limiting or break radius --for each time-- to estimate the radial gradient of the thin disc.

\begin{figure}
\includegraphics[width=0.35\textwidth,angle=-90]{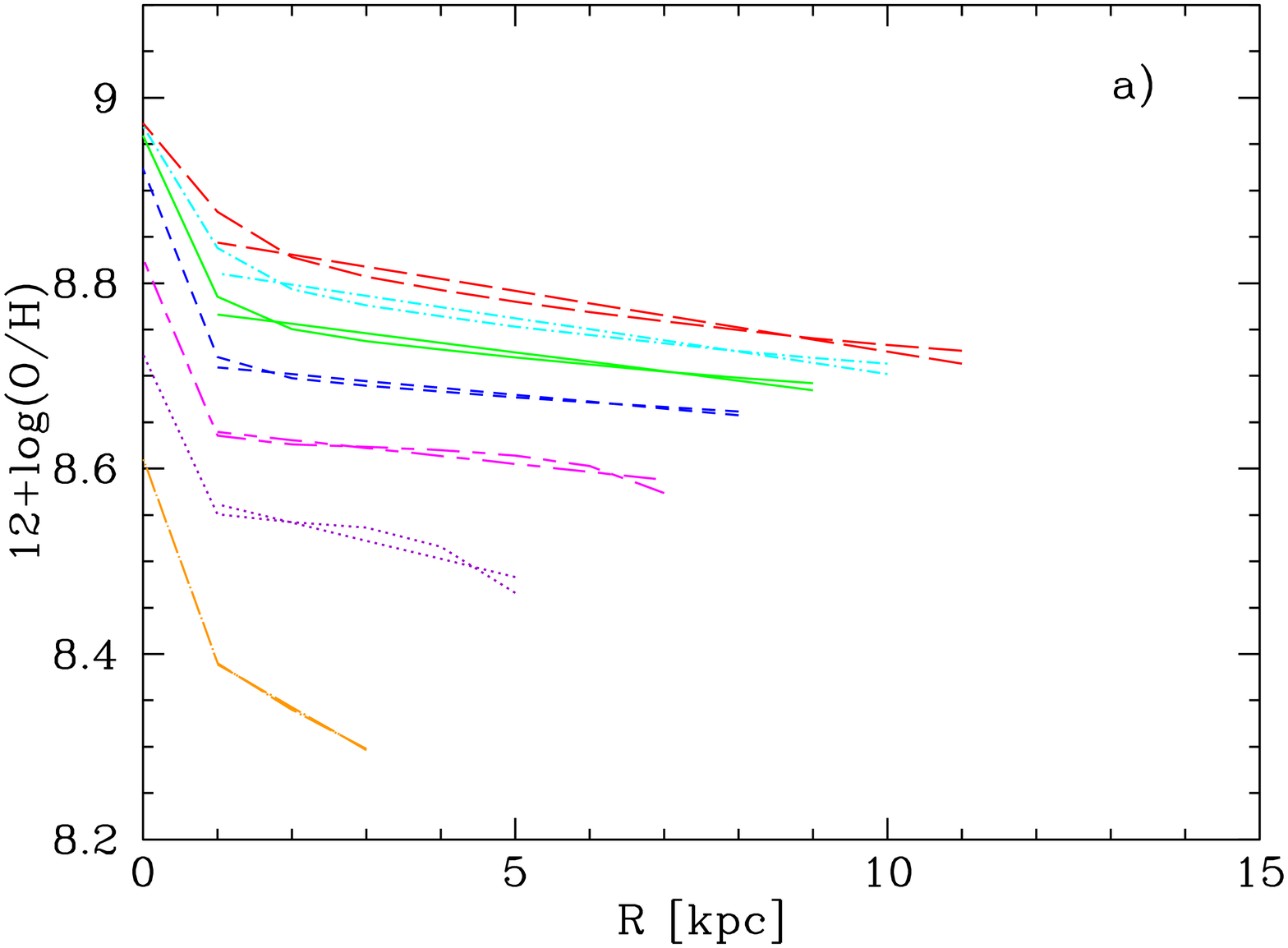}
\includegraphics[width=0.35\textwidth,angle=-90]{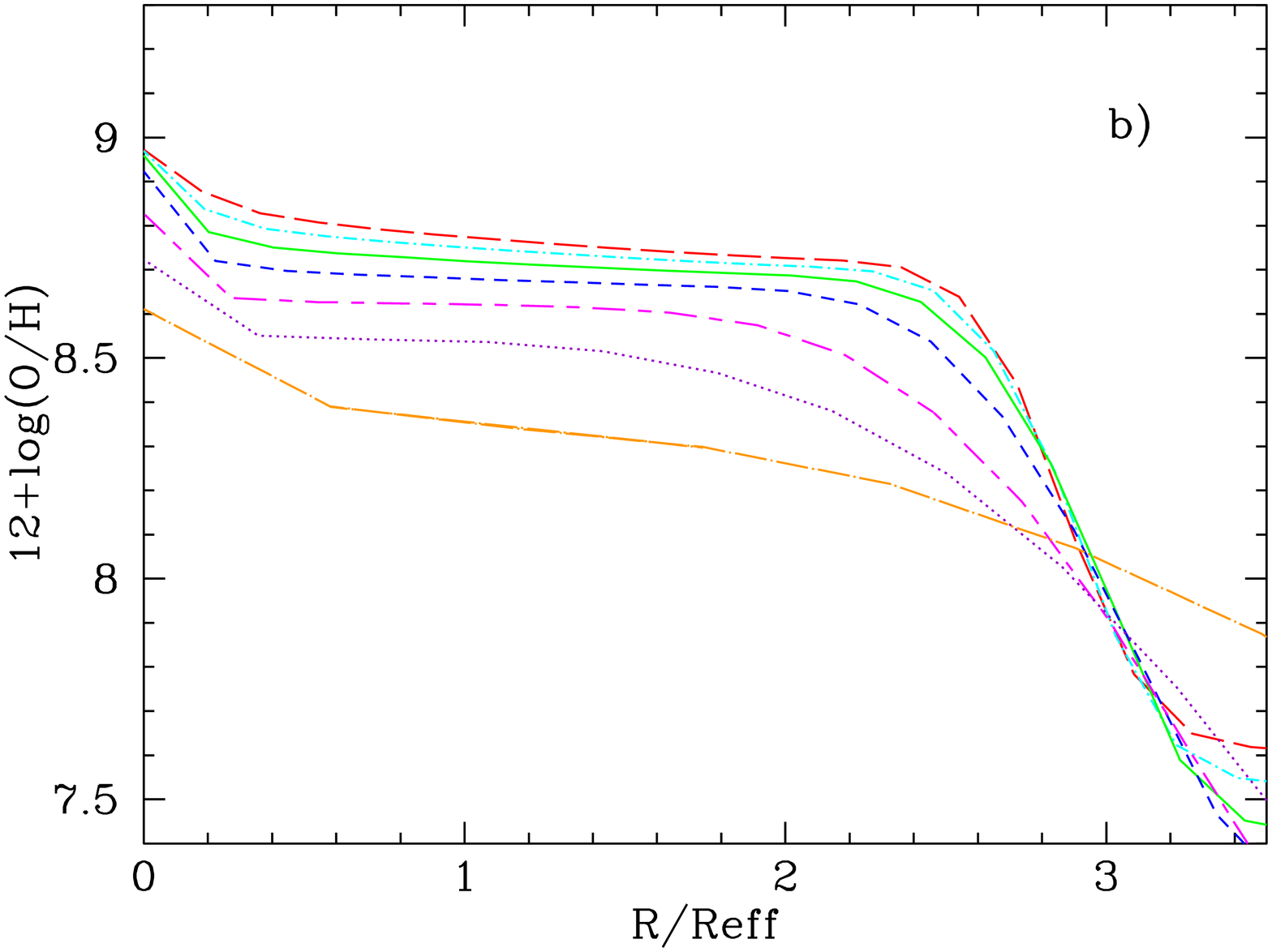}
\caption{The oxygen abundances as $12+\log{({\rm O/H})}$ for MOD1 as a function of: a) the radius $R$, but only for regions located within the break of the radial distributions, $R_{\rm break}$. The least squares straight lines fitting the radial distributions are over-plotted in each redshift; b) the normalized radius $R/R_{\rm eff}$. Line coding is the same as in Fig.~\ref{oh_z}.}
\label{oh_red}
\end{figure}

 For each time step and model calculated, we have computed the effective radius in mass (or half mass radius), defined as the radius in which half of the total stellar mass of the disc (obtained simply by adding the stellar mass of each radial zone and time) is reached. The evolution with redshift of the effective radius, $R_{\rm eff}$, is shown in Fig.~\ref{reff_z} for our 5 models.
The final result, $R_{\rm eff}$(z=0) must be compared with the MWG
data. From the binned observed radial distribution of stellar surface
density for the MWG taken from MOL15, we obtain a $R_{\rm
eff}=5.85$\,kpc, which is in good agreement with the value 5.98\,kpc
given by \citet{dvaupen78}. Moreover, we may find more estimates
for this radius from the measured scale length of the Galactic disc.
If we assume an exponential distribution for the stellar density:
\begin{equation}
\Sigma=\Sigma_{0}\,\exp{-R/R_{\rm D}},
\end{equation}
the total mass of the stars in the disc is $M_{\rm D}=2\,\pi\,\Sigma_{0}\,R_{\rm D}^2$. Using this expression, the relationship of the effective radius with the scale length results:
\begin{equation}
R_{\rm eff}=1.678\,R_{\rm D}.
\end{equation}

\begin{table}
\caption{Observational data for the effective radius $R_{\rm eff}$ of the MWG.}
\begin{tabular}{cl}
\hline
$R_{\rm eff}$ & Reference\\
\hline
5.98 & \citet{dvaupen78}\\
5.95& \citet{vau83} \\
6.56 & \citet{vkruit86}\\
4.49 & \citet{freu96}\\
5.10 & \citet{gould96}\\
4.76 & \citet{zheng01}\\
4.42 & \citet{juric08}\\
4.42 & \citet{sie08}\\
4.42 & \citet{sale10}\\
6.46 & \citet{moni12}\\
3.66 & \citet{bovy13}\\
4.61 & \citet{lic16a}\\
4.42 & \citet{bland16}\\
4.17 & \citet{mcgau16}\\
5.95 & \citet{sofue18}\\
\hline
\hline
\label{re}
\end{tabular}
\end{table}

The photometric scale-length for MWG takes values between $R_{\rm D}=3.86$, and 2.64\,kpc, depending on the technique, author or wavelength band. Therefore, in Fig.~\ref{reff_z} we have used an averaged value, calculated from equation 6 and $R_{\rm D}$ from authors listed in Table \ref{re}, where we give this effective radius $R_{\rm eff}$, in column 1, as obtained from the scale-length $R_{\rm D}$ given by different authors, in column 2. From this method we have obtained an averaged value $<R_{\rm eff}>=5.07\pm 0.93$\,kpc. This value is shown as a purple dot with error bars. We have also plotted the averaged observational point obtained with GAMA data by \citet{lange15} for local disc galaxies, shown as a red triangle at z=0 in the same Fig.~\ref{reff_z}, which is similar to the observational MWG averaged value\footnote{Although this last result implies a slightly larger effective radius at $z=0$ than
we see in our models and than the observed for MWG, this is to be expected since, as \citet{lic16a} explain, our Galaxy seems more
compact than others with the same luminosity. Moreover, most values of  $R_{\rm D}$ from these authors are calculated from the luminosity profile, instead of the mass density radial distribution as we do for the effective radius. So that their results may be slightly different than ours. In fact, the effective radius measured with the mass density profile is $\sim 0.8$ the one measured with luminosity \citep{gon14,gon15}. Therefore, it is reasonable that our results be smaller than the data found  \citep{vdWel14,lange15} for other galaxies.}. All our models end the evolution with a value compatible with the error bars of these observational estimates.
 
The effective radius of all our models follows a similar trend with  redshift. We also included in the same Fig.~\ref{reff_z} the evolution found by \citet[][]{vdWel14} for galaxies with a similar stellar mass than MWG (column corresponding to $\log_{M_{*}}=10.75$ in their Table 2), shown as dashed lines with dots, and the respective $\pm \sigma$ limits, as dotted lines. In this figure we also included the fit given by these authors as a solid black line.  
\citet{xiang18}, studying stellar populations of the local Universe with CALIFA data, also find that the scale-length $R_{\rm D}$ for the radial distribution of surface mass density yields a value of $\sim 4$\,kpc for the young stellar populations and of $\sim 2$\,kpc for the oldest ones. \citet{bland16} have already established that the scale length is smaller for the inner (old populations) regions than for the outer disc (young stellar populations). Taking into account the relationship of $R_{\rm D}$ with the effective radius, this implies that $R_{\rm eff}$ has also been smaller in the past than it is now.

Summarizing, $R_{\rm eff}$ is smaller at $z=4$ than now and the
evolution of the modeled radius is in agreement with observational
estimations. Since the disc is short for higher redshifts, a smaller
radial range should be used to estimate the radial gradient of abundances at other redshifts/times. 

If we assume a Freeman law for disc, the optical radius, defined as
$R_{\rm opt}=R_{25}$ (the isophote of 25 mag/arcsec$^2$ or 
equivalently the radius enclosing the 83\% of the light) is
$R_{\rm opt}\sim 3.2\,R_{\rm D}$, and, using the above relationship
between $R_{\rm D}$ and $R_{\rm eff}$, $R_{\rm opt}\sim $1.9 -- 2.0$\,R_{\rm eff}$, using for both $R_{\rm D}$ and $R_{\rm eff}$ the light profile, (slightly different if the mass density profile is used instead).  As demonstrated by \citet{san14}, all galaxies show a similar behaviour in the radial distribution of oxygen abundances out to 2--2.5\,R$_{\rm eff}$.  Thus, by assuming an effective radius for MWG of $\sim 5.07 \pm 0.93$\,kpc, the radial regions where an uniform radial gradient may be obtained will be within 10.14--12.675 \,kpc. Taking into account the uncertainties in the effective radius, we define an arbitrary break radius as $R_{\rm break}=\sim 2.6\,R_{\rm eff}$, (which gives 13.18\,kpc for the present time). So, our criteria for including only the disc component in the calculation of the gradient is choosing regions located at $R<R_{\rm break}=2.6\,R_{\rm eff}$, which is in good agreement with \citet{san14}. An exact value of this break radius, (2.6, 2.2 or even 2 times the effective radius) can not be given, since the relationship between the optical radius (from the luminosity distribution) and the half mass radius (from the mass distribution) is not actually very well established. The only important thing here is that the same ratio $\rm R_{break}/R_{eff}$ must be used along the time, thus taking the growth of the disk into account for the computation of the radial gradient. This way the break radius will be $\le 4$\,kpc at $z=4$ (just where the radial distribution begins to be steeper, see the orange line in Fig.~\ref{oh_z}) and not 14-15\,kpc as occurs in the present time distribution. Therefore, it is important to  define in each time the radial range over which we can accurately measure the value of the radial gradient in order to estimate its correct evolution.

In Fig.~\ref{oh_red} we show in panel a) a plot similar to
Fig.~\ref{oh_z}, but now using only regions before the break of the
abundance distributions, that is within $R_{\rm break}=2.6\,R_{\rm
eff}$.  These lines will appear slightly steeper for $z=4$ than now,
flattened for $z=3$ and $z=1$, and steepens slightly again for $z=0.5$ until $z=0$. Also, we shown in this Fig.~\ref{oh_red} the least squares straight lines obtained for the disc (eliminating the central bulge, $R=0\,{\rm kpc}$ in the calculation, in order to compute the {\sl disk} radial gradient) which are over-plotted in
each redshift. The resulting radial gradients are given in
Table~\ref{grad_mod}, column 3. The radial gradient is not exactly the same at all times, but, following this approach, the evolution is much smoother than the one from our previous calculations using the whole radial range.
\begin{table}
\caption{Radial gradients of oxygen abundances obtained from MOD1 for different redshifts.}
\begin{center}
\begin{tabular}{cccc}
\hline
Time & z & \multicolumn{2}{c}{$\nabla_{O/H}$}\\
\hline
[Gyr]& & [dex\,kpc$^{-1}$] & [dex\,R$_{\rm eff}^{-1}$]\\
13.20 & 0.0 & $-0.0131 \pm 0.0012$ & $-0.073 \pm 0.007$ \\
10.70 & 0.2 & $-0.0116 \pm 0.0010$ & $-0.061 \pm 0.005$\\
 8.00 & 0.5 & $-0.0106 \pm 0.0010$ & $-0.053 \pm 0.005$\\
 5.50 & 1.0 & $-0.0125 \pm 0.0025$ & $-0.056 \pm 0.011$ \\ 
 2.70 & 2.0 & $-0.0251 \pm 0.0067$ & $-0.092 \pm 0.025$ \\
 1.50 & 3.0 & $-0.0476 \pm 0.0108$ & $-0.133 \pm 0.030 $\\
 0.90 & 4.0 & $-0.0568 \pm 0.0062$ & $-0.098 \pm 0.011$\\
\hline
\end{tabular}
\end{center}
\label{grad_mod}
\end{table}
For $z=4$ the radial gradient is more similar to the one for
$z=0$, now when it is measured out to $R\sim$~3--5\,kpc instead of, say, out to 15\,kpc as in the case for redshift $z$=0. To measure the radial gradient of abundance using a radial range up 14\,kpc when the optical disk has a size not larger than 5\,kpc seems quite unreasonable. To limit the radial range allows a certain and uniform estimate of the radial gradient of the disk and its correct evolution.

Our conclusion is, therefore, that for our models the radial
gradient does not change appreciably with redshift if we measure it
within the thin disc (whose size obviously decreases with increasing
redshift). More specifically, within a break radius $R_{\rm break}=2.6\times R_{\rm eff}$, it flattens from $z=4$ until $z=1$ and steepens again,
although slightly, from $z=1$ until now. In Fig.~\ref{oh_red}, panel
b) we show the same results as in panel a), but now as a function of
the normalized radius $R/R_{\rm eff}$. We see a clear radial gradient
until $R/R_{\rm eff} \sim$~2.3--2.6. Measuring the radial gradient by
fitting least squares straight lines for $R/R_{\rm eff} < 2.6$, we
obtain the values given in column 4 of the Table~\ref{grad_mod}. We
see again that the gradient did not change appreciably since $z =
1.0$.

\subsection{Comparison of models}
\begin{figure*}
\includegraphics[width=0.49\textwidth,angle=0]{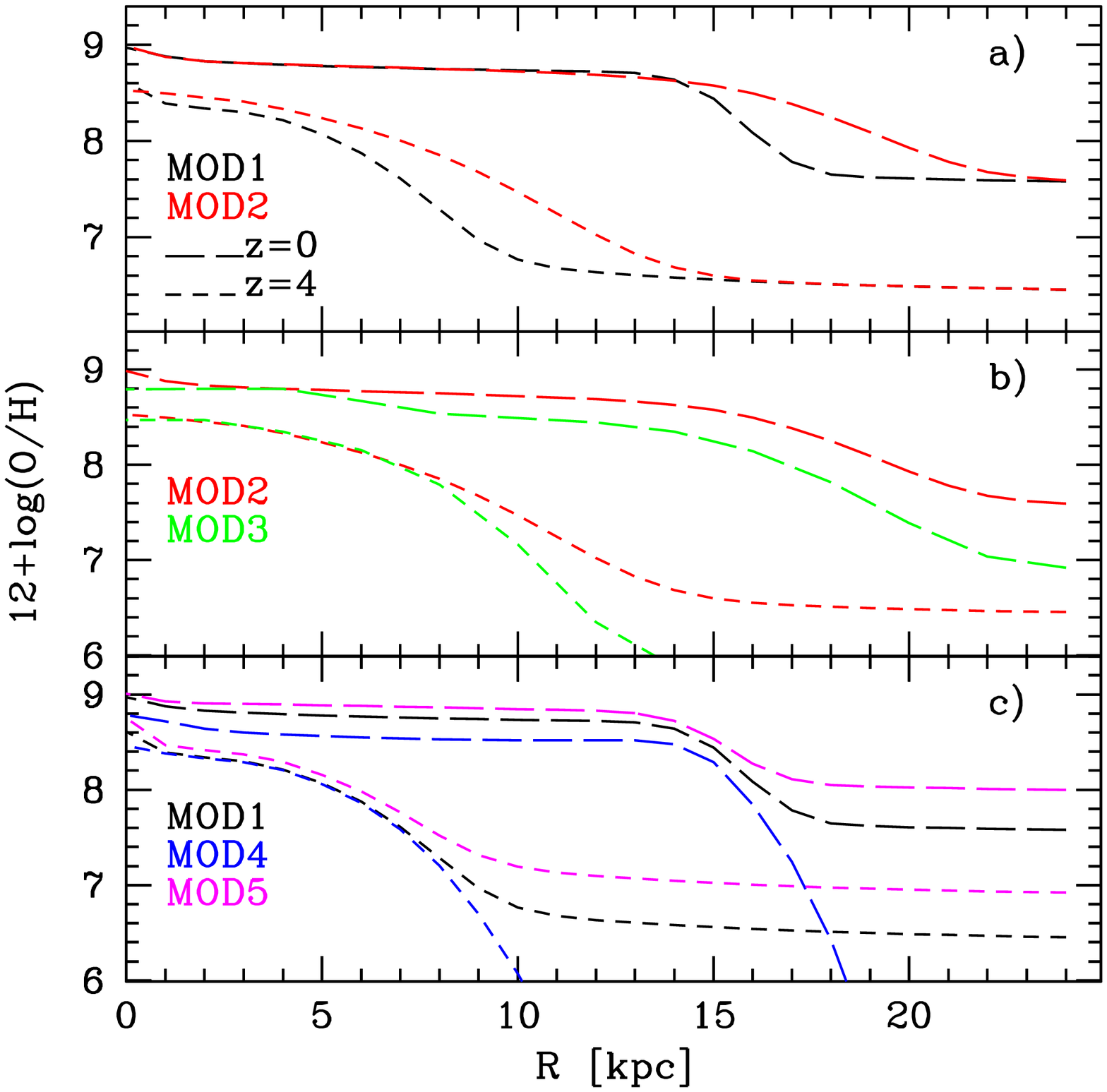}
\includegraphics[width=0.49\textwidth,angle=0]{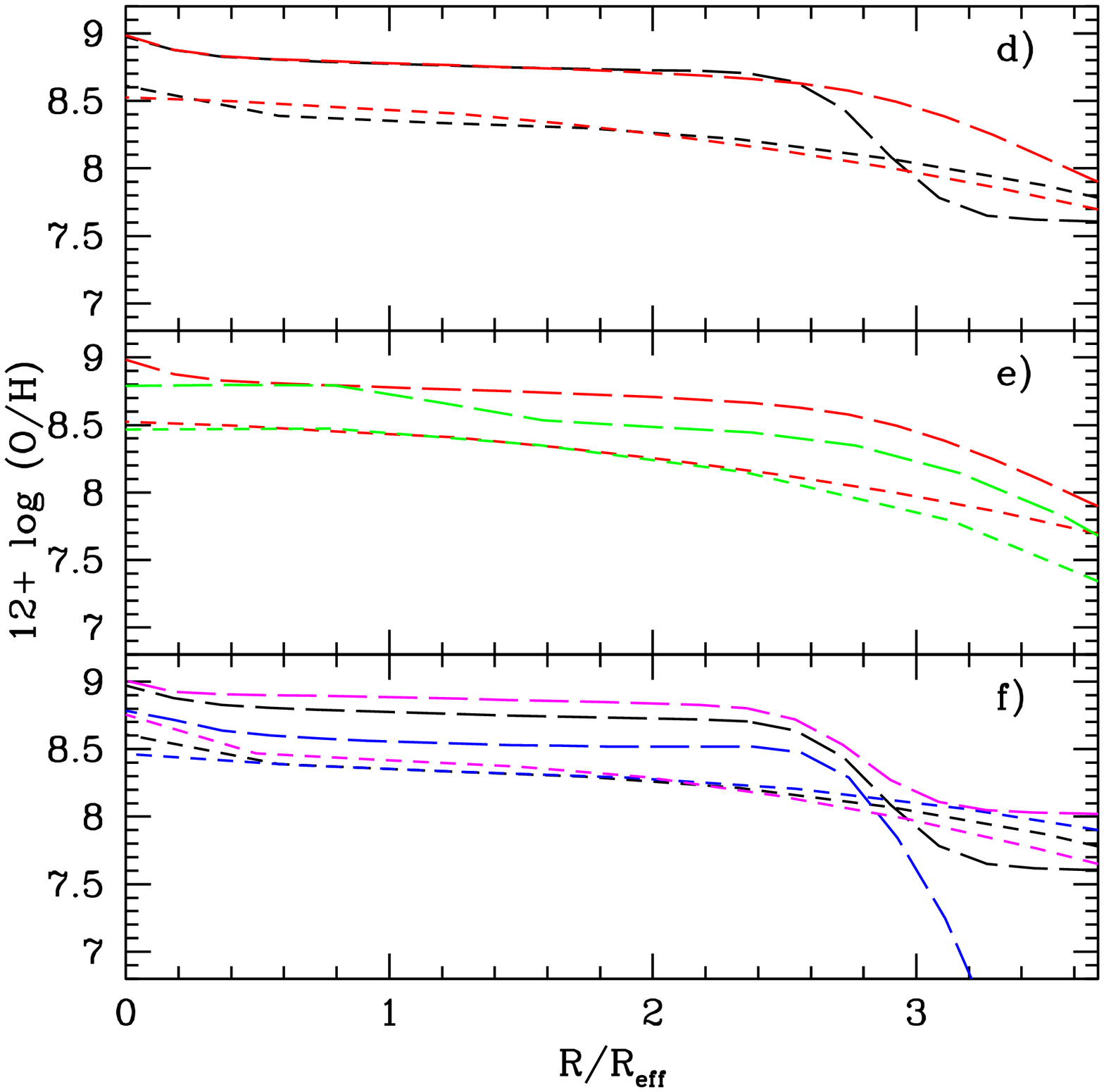}
\caption{The radial distribution of O abundances, $12+\log{({\rm O/H})}$, as a function of the radius $R$ in kpc in the left panels, while in right panels we show the same models as a function of $R/R_{\rm eff}$. Two/three models are shown in each panel: a) and d) MOD 1 and MOD 2, b) and e) MOD2 and MOD3, and c) and f) MOD1, MOD4 and MOD5. Line colors have the same meaning as in Fig.~\ref{reff_z}. Results for redshifts $z=0$ and $z=4$ are shown as long-dashed and short-dashed lines, respectively.}
\label{oh_com}
\end{figure*}

In this subsection we compare MOD1 with the other models listed in
Table~\ref{models}. In the left panels of Fig.~\ref{oh_com} the
comparison is done using $R$, while in the right panels abundances
are represented as $R/R_{\rm eff}$. In the first row of
Fig.~\ref{oh_com}---panels a) and d)---we compare MOD1 and MOD2, where
the infall rate is the same, but the formation of molecular clouds is
done with different prescriptions, as described in \citet{mol17}. We
see that they are clearly different for the outer regions of the disc,
with a smoother radial gradient for MOD2 as compared with MOD1. The
last one shows a strong decrease in the outer disc and then a very
flat radial distribution, in a similar way to the density profile
\citep{wang18}, while MOD2 does not present this strong break, but a
continuously decreasing function. We should note that the effective
radius evolves in a different way in both models, as shown in
Fig.~\ref{reff_z}. Both features balance each other, so that if we see
the distributions as a function of the effective radius, we find a
similar evolution in both cases. Except in the very outer disc, where
differences among models arise, mainly at $z=0$.

We compare in the second row of Fig.~\ref{oh_com} two models using the
same prescription to form molecular clouds, but different infall
rates: MOD2, which uses the new infall rates calculated in MOL16, and
MOD3, which uses the infall rates as in MD05. In fact, both models
also differ in the combination IMF+stellar yields. However, as we
demonstrated in MOL15, all combinations give similar radial gradient
and only absolute abundances are changed by these ingredients. MOD3
has a stronger variation of infall rates with radius than MOD2 and,
consequently, shows a slightly steeper radial gradient of
abundances. Again, however, we find smaller differences when we
compare distributions as a function of $R/R_{\rm eff}$, mainly at
$z=4$.

Another debated question is if the radial gradient maintains the same
slope for all observed radial range or if it flattens (or steepens) in
the outer disc. The radial gradient in MWG might be compatible with a
same slope across the disc until $R\le 18$\,kpc
\citep{alba,est17}. However, for the CALIFA and MUSE surveys, some
authors \citep{san14,sm16,sm18} find a flattening beyond $R> 2--2.5\,R_{\rm
eff}$. We also find a {\sl plateau} with $12+\log{({\rm O/H})}\sim
7.5$ dex for $z=0$ in MOD1 (see Fig.~\ref{oh_z}). In our scenario
there is also star formation in the halo, with an efficiency
$\epsilon_{h}=0.01--0.03$, as chosen in \citet{fer92} to reproduce the star
formation history and age-metallicity relation in the Galactic halo,
and used in MOD1, MOD2 and MOD3. This implies that the gas
infalling in the disc is not primordial but it is pre-enriched,
contributing to the level of metal abundances in the disc. Its effect
is not very apparent in the bright regions of the disc, with much
higher abundances than the ones of the infalling gas, but it may be
seen clearly in the outer regions, where the star formation rates are
very low and, consequently, the infall may enrich the ISM of these
regions.  In order to check if this possibility may produce a
flattening of the oxygen abundances in the outer disc, we have now
calculated another two models, both similar to MOD1 but modifying the
star formation efficiency of the halo to be $\epsilon_{h}=0.00$ (primordial abundances for the infalling gas) and $\epsilon_{h}=0.10$ in MOD4 and
MOD5, respectively, in order to check if the outer regions of the disc
are modified as consequence of the different infall enrichment. 
\begin{figure*}
\includegraphics[width=0.35\textwidth,angle=-90]{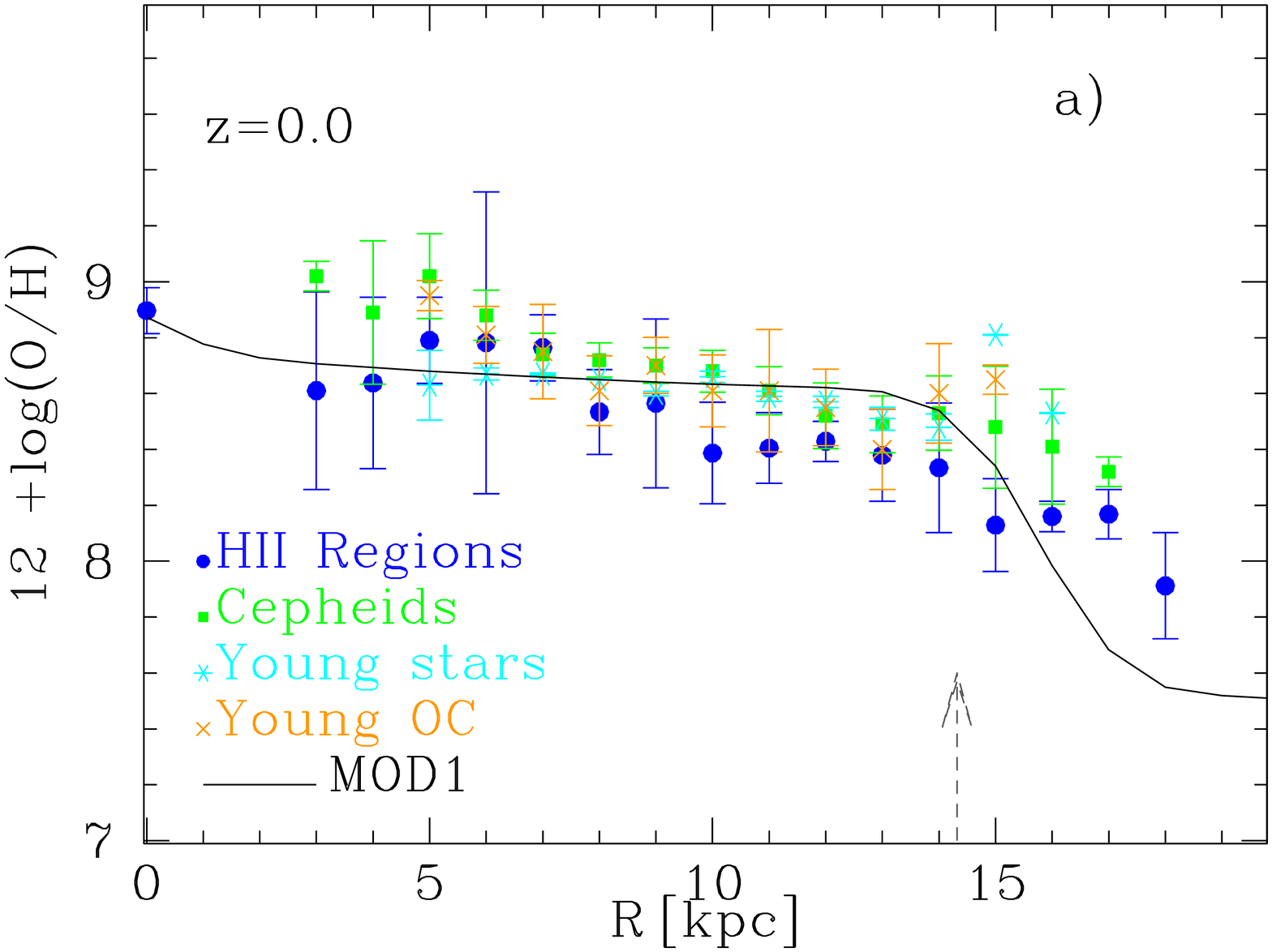}
\includegraphics[width=0.35\textwidth,angle=-90]{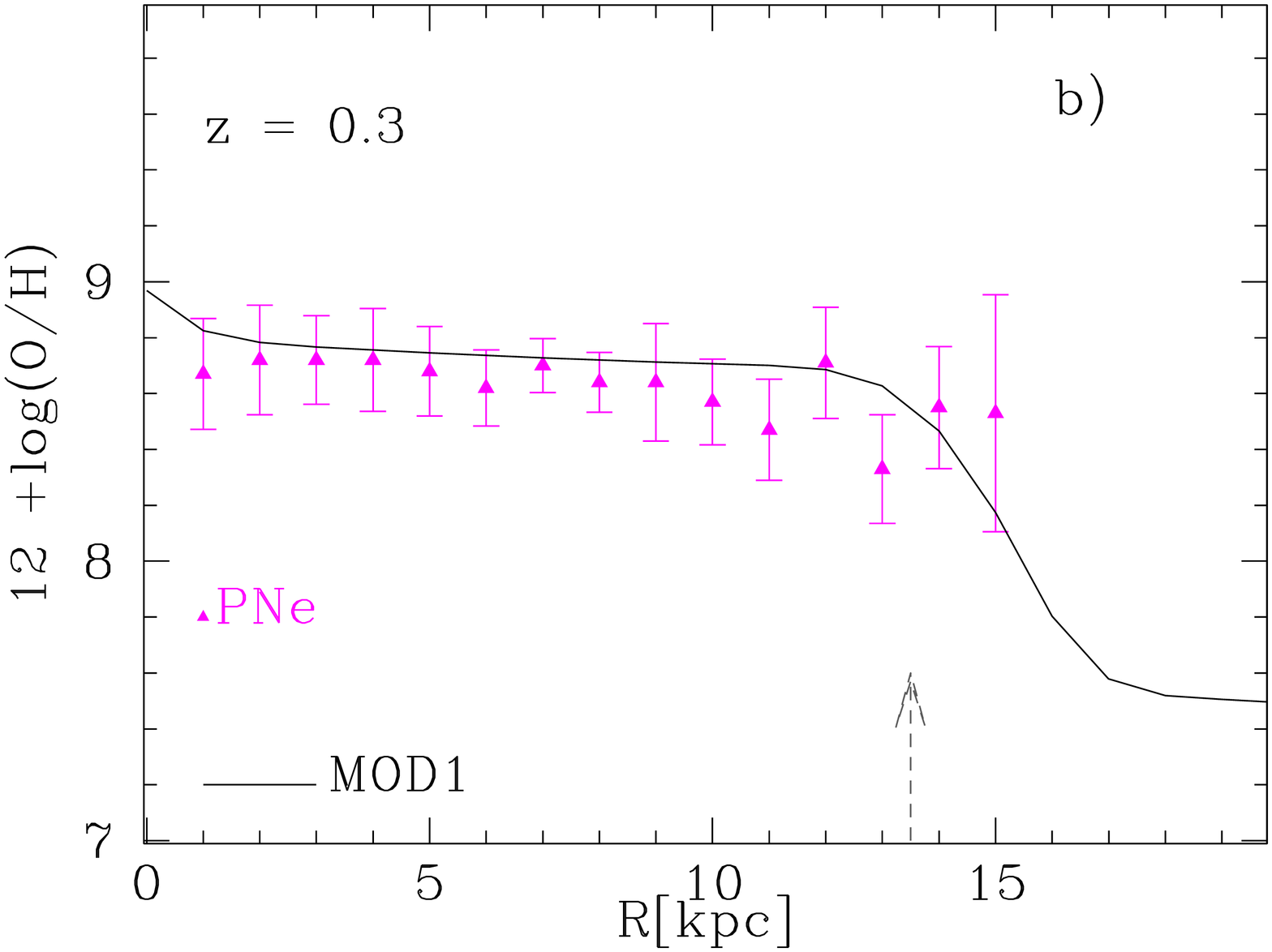}
\includegraphics[width=0.35\textwidth,angle=-90]{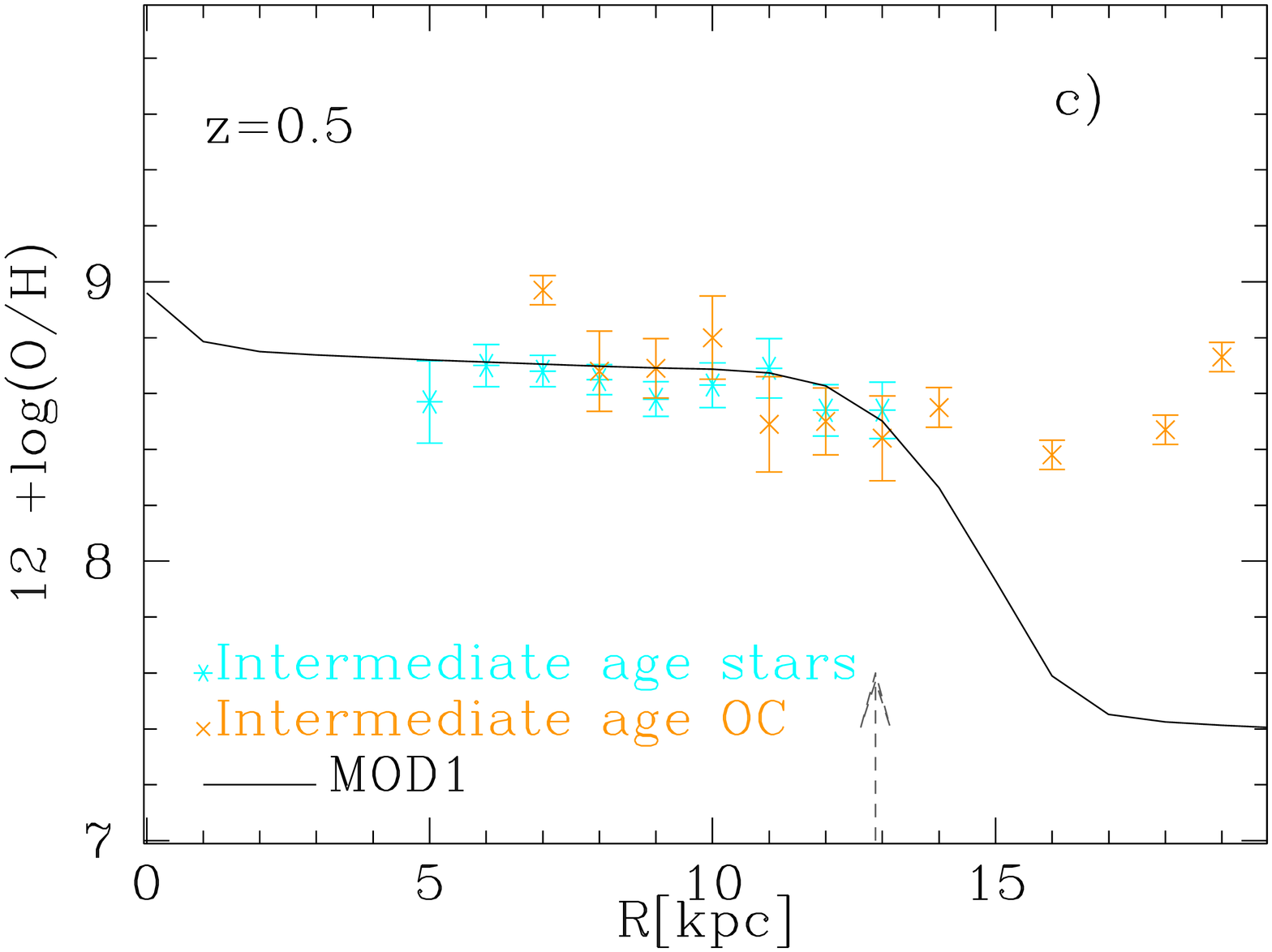}
\includegraphics[width=0.35\textwidth,angle=-90]{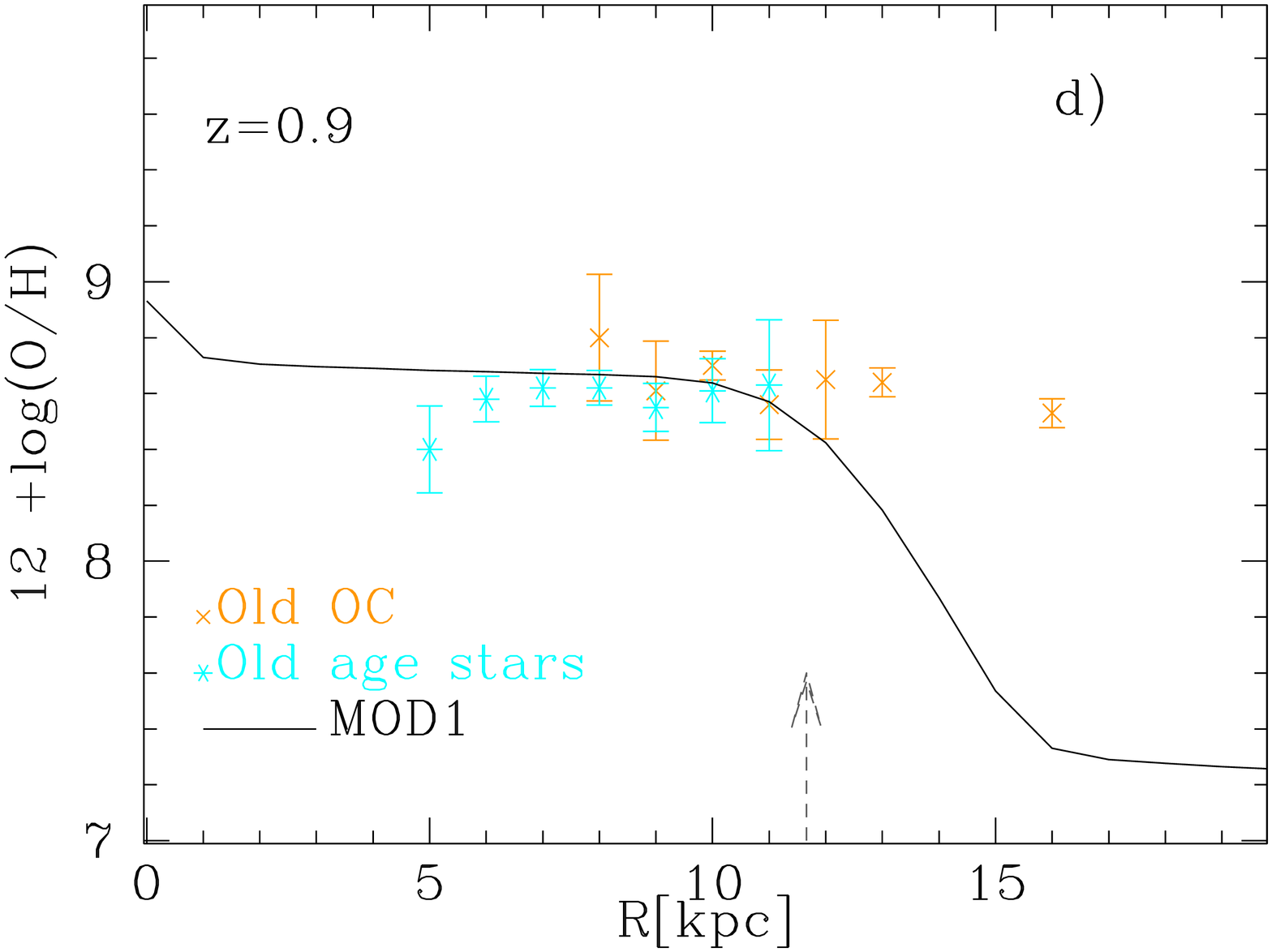}
\caption{The oxygen abundances as a function of the galactocentric radius $R$. Panels show the comparison of MOD1: a) at the present time ($z=0$) with data from H{\sc ii} regions, Cepheid stars, young stars and young OC, as blue circles, green squares, cyan  stars and orange crosses, respectively; b) at $z=0.3$ and PNe data as magenta triangles; c) at $z=0.5$ with intermediate age stars and OC, as cyan stars and orange crosses, respectively; d) at $z=0.9$ with data from old OC and old stars, as orange crosses and cyan stars, respectively. In each panel an arrow indicates the size of the expected break, $R_{\rm break}\rm \sim 2.60\times \rm R_{eff}$. All points are the binned results obtained as explained in Section ~\ref{data}.} 
\label{oh_obs_mod}
\end{figure*}
We see in panel c) of Fig.~\ref{oh_com} that the oxygen abundance for
$z=0$ in the external disc drops continuously in MOD4, while it takes
a constant value of $\sim 8.0$\,dex for MOD5. This value is $\sim
0.5$\,dex higher than in MOD1, where this outer abundance is $\sim
7.5$ dex.  Therefore, a flattening of the gradient in the outer
discs, with a value of $12+\log({\rm O/H}) \sim 8.0$ dex as observed by
CALIFA and MUSE, is in agreement with this possibility of star
formation in the halo, and the subsequent metal enriched infall. This
hypothesis of an enriched infall as the cause of the flat radial
distributions of abundances in the outer disc was also proposed by other
authors, as \citet{bresolin12,bresolin16}. This abundance will be
similar for all galaxies when a same efficiency for the star formation
in the halo is used. A different plateau for each galaxy would be also
reasonable if the halo has a different evolution in each case,
depending on the environment or on the interactions suffered by the
proto-halo in a early phase of the evolution. We should note that in
the literature there are other interpretations for these flattening of
the gradient at the outer disc as, e.g., a star formation efficiency
that is roughly constant for $R > R_{\rm eff}$ \citep[see][ for
details]{kudritzki15}.

In any case, when we represent these distributions as a function of a
normalized radius (right panels of Fig.\ref{oh_com}), we found that all
of models show a very similar radial gradient, with only small
differences among them. This has previously been shown for a suite of
simulated disc galaxies in \cite{few12}, where while the metallicity
gradients with normalized radius have a wider spread than seen here,
no trends are found with mass or environment (in a limited sense) as
are found in absolute gradient \citep[this analysis is expanded
in][]{pil12}. We find here that the star formation, the infall rate or
the stellar yields do not modify the basic value of gradient. This
gradient shows a very similar slope for all redshifts, particularly if
we use only the disc regions, without the central region at $R=0$,
where we represent the bulge.

\section{Discussion: Evolution of the radial gradient}

\subsection{Evolution of the gradient with time}

In the time range $3 \le t \le 13.2$\,Gyr, the possible evolution of
the gradient seems to be rather undetectable. Thus, a clear separation in data coming from H\,{\sc ii} regions and PNe of 2--4\,Gyr old will not
be apparent, such as SH10, MAC13, and M16 claim. We see this clearly
in Fig.~\ref{oh_obs_mod}, where O abundances of MOD1 are compared in:
a) for $z=0$ with our binned data from H{\sc ii} regions, Cepheid
stars, young stars and young OC; in panel b) for $z=0.3$ with PNe data; in panel c) for $z=0.5$ with intermediate age stars and OC data; and in panel d) for $z=0.9$ with old age stars and OC.

The model results agree with the generic trend of observational data
in all panels of Fig.~\ref{oh_obs_mod}, although there are dots
corresponding to OC data slightly above the line in panels c) and d) for the outer regions represented. 
This result may be related to previous arguments given in Subsection \ref{sec:model_descrition}: in the outer regions, data
would not correspond to the thin disc component, but to the thick
disc or the halo populations. For $z=0$ the optical radius for the MWG is around 11\,kpc and the drop begins at $R=14$\,kpc. For $z=0.9$, following our models, it would be $R_{\rm opt} \sim 9$\,kpc, and the first break would be at $R\sim 11$\,kpc, as shown in MOD1. It is, therefore, possible that last points after $R_{\rm break}$ pertain to the transition region thin-thick disc, before to reach the real outer region for which a flatter radial gradient is
expected. There are some studies \citep{nord04,all06,chen12,car12,boe14} using data from the thick disc which also found flat or even positive radial gradient for the largest galactocentric distances. 
These data in the outer regions may also be the consequence of a stellar radial migration process, meaning that old objects are located further from their birth radius.

\begin{figure}
\includegraphics[width=0.35\textwidth,angle=-90]{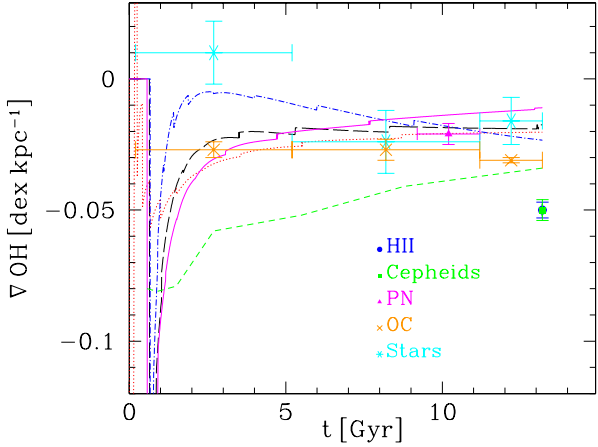}
\caption{Evolution of the O radial gradient along time for different objects. The predictions of models are shown with the same coding of lines than in Fig.6.}
\label{grad_t}
\end{figure}

The time evolution of the gradient is shown in Fig.~\ref{grad_t}. In
this figure, we represent the corresponding radial gradient obtained
in Table 2, in the times associated, following the assumed age, to
each objects: H{\sc ii} regions and Cepheids at $t=13.2$\,Gyr; Young
stars and OC between $t=11$ and the present time; PNe between 9 and
11\,Gyr; intermediate age stars and OC between 5 and 11\,Gyr; and old
stars and OC between 0 and 5\,Gyr. Over them, we plot the gradients
obtained by our 5 models. When the gradients are measured within the
corresponding disc for each time, most of them do not give a strong evolution in time for the O radial gradient as it is shown, mainly for MOD1, MOD2 and MOD5; indeed they are around
$-0.02$ or $-0.03$\,dex\,kpc$^{-1}$ from $t=4$\,Gyr until now. MOD3 (with the old prescription of infall of gas from MD05) clearly does not fit this evolution shown by data, while MOD4 show a
very strong flattening from $t=1$\,Gyr until $t=2$\,Gyr, and a continuous steepening since then until now.

If we accept this behavior of our models, the radial gradient of O
abundances do not evolve very much with time in the last 10\,Gyr. The
observational data presented in Fig.~\ref{grad_t} also supports this
conclusion. Following these data the radial gradient has been
basically the same for times from 4-5\,Gyr until 13\,Gyr, only showing
a steepening in the last times. Only the flat radial distribution for
the oldest stars is out of this trend, showing a large difference with the point of the oldest OC, (although it is necessary to take into account that the radial range of the old stars is very narrow compared with the one in which other object abundances have been measured). As we demonstrate above, finding this smooth behavior for the slope, basically without or with a mild evolution, is mainly the consequence of restricting the radial range in our models to the break radii, which is around $R_{\rm break}=2.6\times R_{\rm eff}$. In the case of the observational data, this result implies that objects beyond this limit might be members of the thick disc or the halo. Other possibility is that some stars or OC have migrated to the outer disc. This would be more likely for the old objects, which would explain the points above the line of the model in Fig.9c and Fig.9d.  It would imply that these objects do not represented the abundances of the gas of the regions were are located, but the ones of the regions were they were created. Curiously, this seems to occur more for old OC than for old stars.

Our conclusion is that, in any case, it is necessary to take into account that the disc grows with time, and that at the time when objects with old/intermediate ages were created, the disc was smaller than now (see Section 3). Therefore, if we want to measure with good accuracy the evolution of the radial gradient of abundances in the thin disc, we need to determine very carefully which component are we observing and the radial range where to measure it.

\subsection{Evolution of the gradient with redshift}
\label{zt}

\begin{figure}
\includegraphics[width=0.35\textwidth,angle=-90]{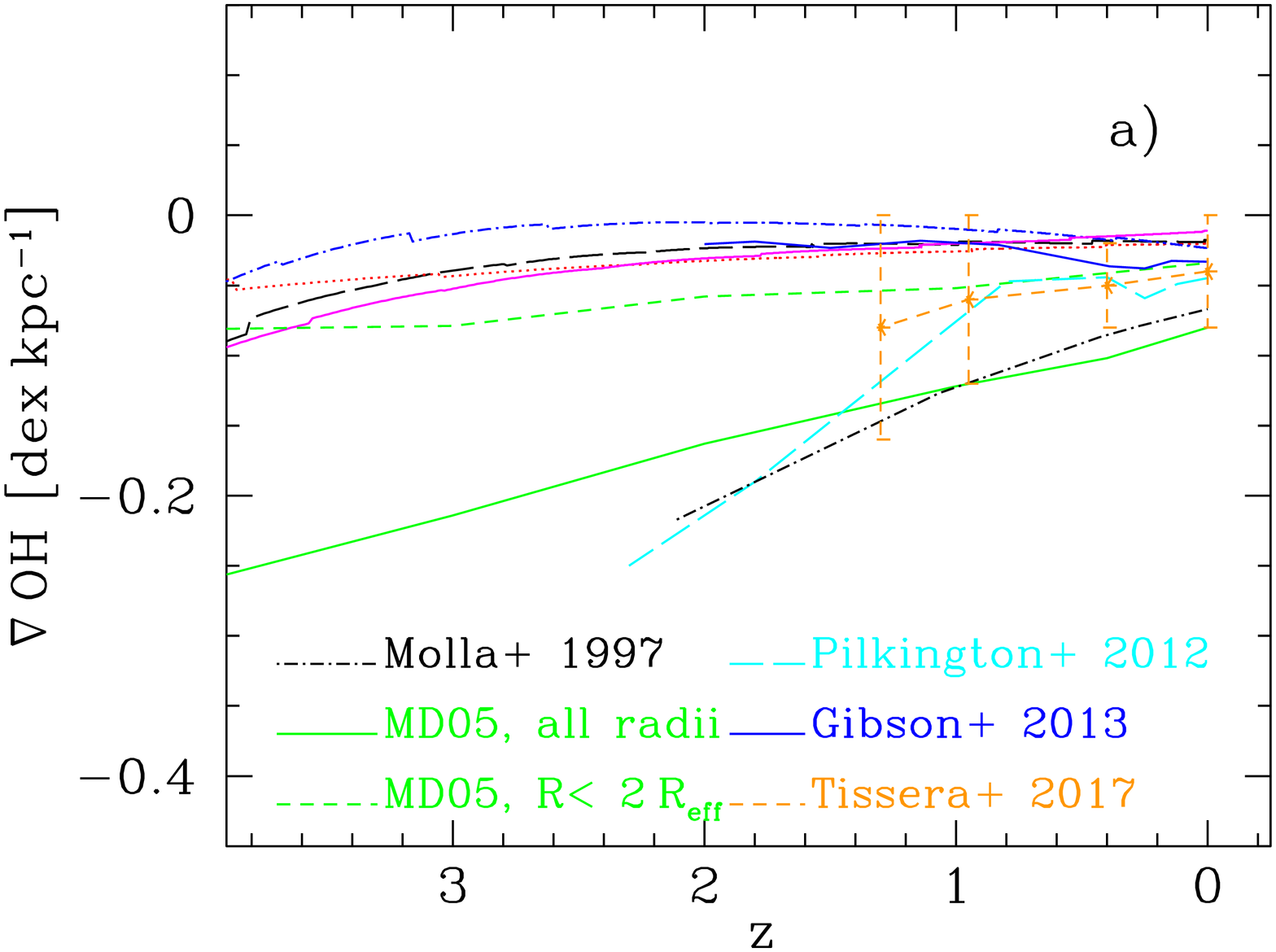}
\includegraphics[width=0.35\textwidth,angle=-90]{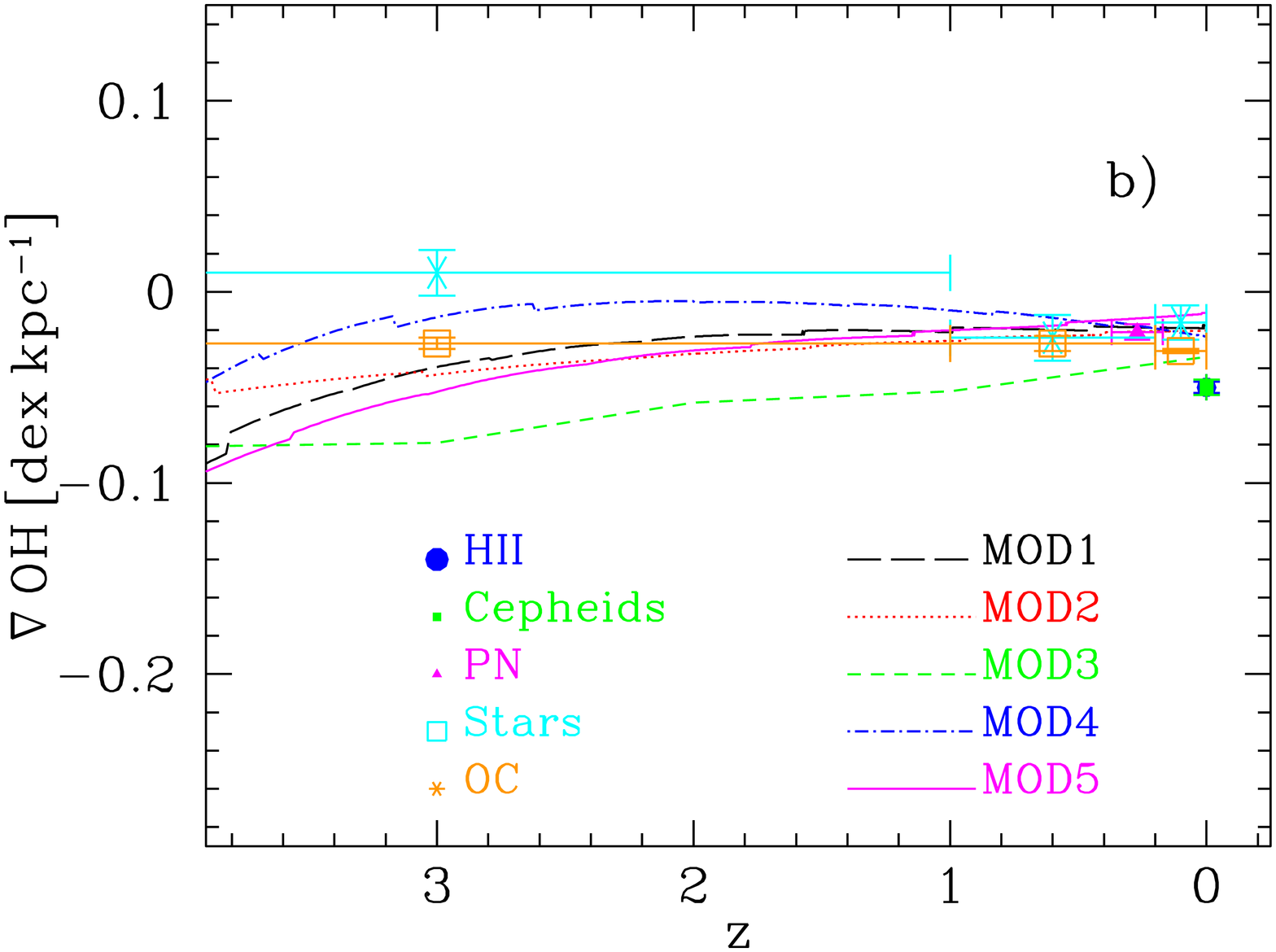}
\includegraphics[width=0.35\textwidth,angle=-90]{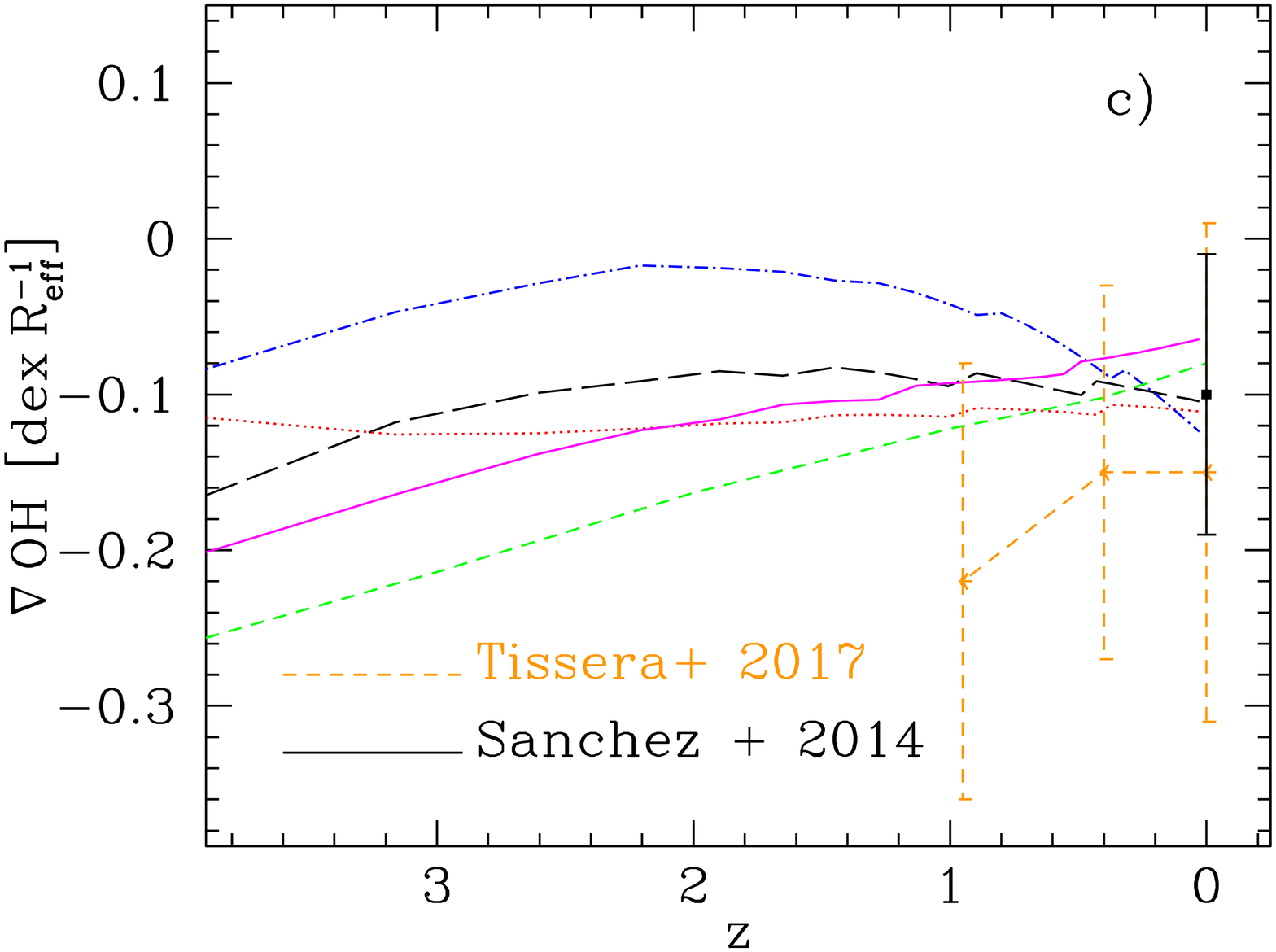}
\caption{The  evolution with redshift of the radial gradient of O abundances: a) in dex\,kpc$^{-1}$ units compared with cosmological simulations from \citet{gib13} and \citet{tis17}, as labelled; b) compared with observational data from different objects compiled in this work, as labelled; c) in dex$\,R_{\rm eff}^{-1}$ units compared with cosmological simulations from \citet{tis17} and data from \citet{san14,sm16} for the present time. All panels show models MOD1 to MOD5 with the same line coding as labelled in panel b).}
\label{grad_z_mwg}
\end{figure}

In other works related with the evolution of the radial gradient of
metallicity, data or model results are shown as a function of
redshift. For that reason, in order to compare the predictions of our
models with other results in the literature, we shown here the
predictions for earlier times as a function of the redshift, instead
of time. We show in Fig.~\ref{grad_z_mwg} the evolution of the
gradient, $\nabla_{OH}$, along redshift for our models.  In panel a)
the results are compared with previous chemical evolution models
\citep[][ and MD05 --here MOD3]{mol97} and cosmological simulations
from \citet{pil12,gib13,tis17} as labelled. \citet{pil12} gave results
similar to MD05 and \citet{mol97} using all radial regions. The recent
results from \citet{tis17} are closer to MD05 (MOD3) computed for
regions within the optical disc, while MOD1, MOD2 and MOD5 give a
behavior more similar to \citet{gib13}, with a smooth evolution almost
nonexistent for some of them. MOD4 (without SFR in the halo) show a
clear flattening until $z=1$ and then a steepening in the last times,
while in MOD3 the gradient is continuously flattening as MOD5 also
does. MOD1 and MOD2 have a similar shape as MOD4 but smoother than
this one.

In panel b) of Fig.\ref{grad_z_mwg} we compare our results with the
same data as in Fig.~\ref{grad_t}. Although our models show a smooth
behavior for the gradient in agreement with these observational
results, we see that the gradient was steeper at higher redshifts
($z> 2$ or 3). This is important because it is related to the phase
of the disc formation. At a time of $t=1 \, {\rm or } \, 2$\,Gyr, 
when there is sufficient gas in the centre of the future disc,
star formation begins and the first elements appears in the
ISM. This is the way in which we achieve higher abundances in the
central regions while the surroundings remain almost primordial, 
resulting in a strong radial gradient of abundances. These first
phases are numerical unstable (as it is seen in Fig.~\ref{grad_t}), making it more difficult to measure the gradient. In fact, real galaxies also will have unstable phases
and this fact must be taken into account when observations will be
performed over discs probably in the formation process.

In panel c), we represent $\nabla_{OH}$ {\sl vs.} the normalized
radius ${R/R_{\rm eff}}$. MOD1, MOD2, even MOD5, are within the
evolution obtained by \citet{tis17}, albeit in the shallower
frontier. Our resulting gradient is in very good agreement with the
universal value from \citet{san14,sm16,sm18} at $z=0$ for all
models. In this case the evolution seems stronger than in previous
panels, due to the evolution of the effective radius. Now differences
between models are larger: MOD3 and MOD5 show a continuous flattening,
MOD2 is basically flat and MOD1 and MOD4 show a flattening until
$z\sim 1$ with a steepening at the end.

\section{CONCLUSIONS}
In this work we have highlighted the importance of determining the radial gradient of O abundances in the disc of the Milky Way Galaxy. We summarize the most important results obtained in this work.
\begin{enumerate}
\item  It is essential to determine the disc radial gradient always within the optical radius or a radius slightly higher, (we use a maximum  value  $R_{\rm break} \le 2.6\,R_{\rm eff}= 1.3\,R_{\rm opt}$ for calculating it). We claim that the use of a variable radial range, which takes into account the growth of the disk along the time or redshift, is an important caveat for estimating the correct evolution of the {\sl disk} radial gradient.

\item We built five models to analyse the time evolution of the O radial gradient. The influence of the infall of gas was examined in model MOD3, which uses an infall prescription from \citet{md05}. The others four models use the new prescription from \citet{mol16} which give smoother infall of gas. We also tested the influence of the prescriptions of H{\sc i} to H$_{2}$ conversion process through comparison of models MOD1 \citep[prescription named ASC in ][]{mol17} and MOD2 \citep[prescription named STD in][]{mol17}. We tested the hypothesis of having an enriched infall from the halo, by using different star formation efficiency in the halo, a higher efficiency in MOD5 than in the other models and the hypothesis of no star formation in the halo (primordial abundances for the infalling gas) in MOD4.

\item The evolution of the gradient measured within the thin disc until a $R_{\rm break}$ variable with time is mild for MWG, with an average value $\nabla_{\rm OH} \sim -0.02 \, {\rm or} \, -0.03 \rm \,dex\,kpc^{-1}$ with slight differences among models.

\item When it is measured as a function of normalized radius, we
obtain a value $\nabla_{OH} \sim -0.10 \,{\rm dex}\,R_{\rm eff}^{-1}$
for $0<z<1.5$ in models MOD1, MOD2 and MOD4 with our new infall prescription from \citet{mol16}, which is in excellent agreement with the
local universal gradient found by \citet{san14,sm16,sm17} for
CALIFA and MUSE surveys galaxies. Taking into account that we
calibrate our model to reproduce the radial distributions of the disc
(gas, stars, SFR and elemental abundances), but not the effective
radius, nor the normalized gradient, we consider this fit a success of
our model. The gradient is slightly flatter for models MOD3 (infall type from \citet{md05}) and MOD5 (lower star formation efficiency in the halo).
MOD4 (no star formation in the halo) shows an early flattening and then a steepening, reaching in the
end the same present day value as observed.

\item The oxygen abundance for $z=0$ in the outer regions drops continuously in MOD4, while it takes a constant value of $\sim 8.0$\,dex for MOD5. This value is $\sim 0.5$\,dex higher than in MOD1, where this outer abundance is $\sim 7.5$ dex. Therefore, a flattening of the gradient in the outer discs, with a value of $12+\log({\rm O/H}) \sim 8.0$ dex as observed by CALIFA and MUSE, is in agreement with the possibility of star formation in the halo, and the subsequent metal enriched infall.

\item The radial gradient of abundance in isolated galaxies is a scale
effect which depends on disc growth. When measured within the optical
disc, it maintains a similar value as it evolves where this occurs
quietly i.e., without interactions with the environment and once the
disc is formed, at least until $z=1.5$. For higher redshifts, our
models have different evolutions, depending on the hypothesis used for
the infall rate, to form molecular clouds, or to form stars in the
halo. The stronger radial gradient appears in the early phase of the
disc formation, defined by the dynamical mass of each
galaxy. This point will be analysed in more detail in the next future, when models for other type/mass of galaxies and the corresponding
radial gradients will be presented.

\end{enumerate}

These conclusions highlight the needed for further investigation in
order to clarify the differences between theoretical models, as well
as to extract out and interpret observations at intermediate and high
redshifts.  We will address in a forthcoming paper the possible variations
of these results for other spiral galaxies, analyzing the dependence
on the stellar mass and size (Moll{\'a} et al. in preparation).

\section*{Acknowledgements}
The authors acknowledge the anonymous referee for his/her helpful comments.
This work has been partially supported by MINECO-FEDER-grants
AYA2013-47742-C4-4-P, AYA2016-79724-C4-1-P and AYA2016-79724-C4-3-P.
YA is supported by contract RyC-2011-09461 of the \emph{Ram\'on y
Cajal} programme. BKG \& CGF acknowledge the support of STFC through
the University of Hull Consolidated Grant ST/R000840/1, and access to
\sc viper\rm, the University of Hull High Performance Computing
Facility. This research was supported in part by the National Science
Foundation under Grant No. PHY-1430152 (JINA Center for the Evolution
of the Elements). WJM acknowledges support of FAPESP process 2018/04562-7.

\end{document}